\documentclass[twoside,twocolumn,9pt]{article}

\usepackage{extsizes}
\usepackage[super,sort&compress,comma]{natbib} 
\usepackage[version=3]{mhchem}
\usepackage[left=1.5cm, right=1.5cm, top=1.785cm, bottom=2.0cm]{geometry}
\usepackage{balance}
\usepackage{mathptmx}
\usepackage{sectsty}
\usepackage{graphicx} 
\usepackage{lastpage}
\usepackage[format=plain,justification=justified,singlelinecheck=false,font={stretch=1.125,small,sf},labelfont=bf,labelsep=space]{caption}
\usepackage{float}
\usepackage{fancyhdr}
\usepackage{fnpos}
\usepackage[english]{babel}
\addto{\captionsenglish}{%
  
}
\usepackage{array}
\usepackage{droidsans}
\usepackage{charter}
\usepackage[T1]{fontenc}
\usepackage[usenames,dvipsnames]{xcolor}
\usepackage{setspace}
\usepackage[compact]{titlesec}
\usepackage{hyperref}

\usepackage{epstopdf}

\usepackage{cleveref} 
\usepackage{physics}    
\usepackage{bm}
\usepackage{comment}

\newcommand{\ee}{\mathrm{e}}

\definecolor{cream}{RGB}{222,217,201}

\begin{document}

\pagestyle{fancy}
\thispagestyle{plain}
\fancypagestyle{plain}{
\renewcommand{\headrulewidth}{0pt}
}

\makeFNbottom
\makeatletter
\renewcommand\LARGE{\@setfontsize\LARGE{15pt}{17}}
\renewcommand\Large{\@setfontsize\Large{12pt}{14}}
\renewcommand\large{\@setfontsize\large{10pt}{12}}
\renewcommand\footnotesize{\@setfontsize\footnotesize{7pt}{10}}
\makeatother

\renewcommand{\thefootnote}{\fnsymbol{footnote}}
\renewcommand\footnoterule{\vspace*{1pt}%
\color{cream}\hrule width 3.5in height 0.4pt \color{black}\vspace*{5pt}} 
\setcounter{secnumdepth}{5}

\makeatletter 
\renewcommand\@biblabel[1]{#1}            
\renewcommand\@makefntext[1]%
{\noindent\makebox[0pt][r]{\@thefnmark\,}#1}
\makeatother 
\renewcommand{\figurename}{\small{Fig.}~}
\sectionfont{\sffamily\Large}
\subsectionfont{\normalsize}
\subsubsectionfont{\bf}
\setstretch{1.125} 
\setlength{\skip\footins}{0.8cm}
\setlength{\footnotesep}{0.25cm}
\setlength{\jot}{10pt}
\titlespacing*{\section}{0pt}{4pt}{4pt}
\titlespacing*{\subsection}{0pt}{15pt}{1pt}

\fancyfoot{}
\fancyfoot[RO]{\footnotesize{\sffamily{1--\pageref{LastPage} ~\textbar  \hspace{2pt}\thepage}}}
\fancyfoot[LE]{\footnotesize{\sffamily{\thepage~\textbar\hspace{3.45cm} 1--\pageref{LastPage}}}}
\fancyhead{}
\renewcommand{\headrulewidth}{0pt} 
\renewcommand{\footrulewidth}{0pt}
\setlength{\arrayrulewidth}{1pt}
\setlength{\columnsep}{6.5mm}
\setlength\bibsep{1pt}

\makeatletter 
\newlength{\figrulesep} 
\setlength{\figrulesep}{0.5\textfloatsep} 

\newcommand{\topfigrule}{\vspace*{-1pt}%
\noindent{\color{cream}\rule[-\figrulesep]{\columnwidth}{1.5pt}} }

\newcommand{\botfigrule}{\vspace*{-2pt}%
\noindent{\color{cream}\rule[\figrulesep]{\columnwidth}{1.5pt}} }

\newcommand{\dblfigrule}{\vspace*{-1pt}%
\noindent{\color{cream}\rule[-\figrulesep]{\textwidth}{1.5pt}} }

\makeatother

\twocolumn[
  \begin{@twocolumnfalse}
{\includegraphics[height=30pt]{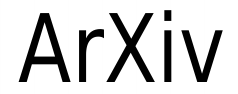}\hfill\raisebox{0pt}[0pt][0pt]{\includegraphics[height=55pt]{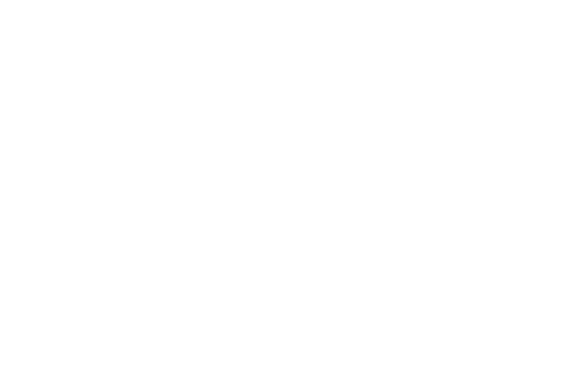}}\\[1ex]\includegraphics[width=18.5cm]{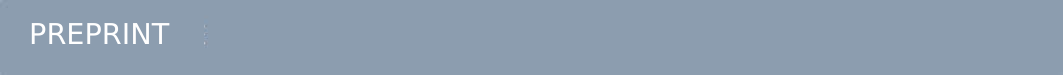}}\par
\vspace{1em}
\sffamily
\begin{tabular}{m{4.5cm} p{13.5cm} }

& 
\noindent\LARGE{\textbf{Dipolar Attraction of Superparamagnetic Nanoparticles$^\dag$}} \\
\vspace{0.3cm} & \vspace{0.3cm} \\

 & \noindent\large{Frederik Laust Durhuus,\textit{$^{a,\ddag}$} Marco Beleggia\textit{$^{b,c}$}, Cathrine Frandsen\textit{$^{a,\S}$}} \\

& \noindent\normalsize{
For superparamagnetic nanoparticles (SMNPs), it is often claimed that the rapid thermal fluctuations of their magnetic moments negates the magnetic dipolar attraction, hence preventing aggregation in liquid suspension. However we find that this is a misconception. 
Using Langevin dynamics, we simulate SMNP pairs and the dimer clusters they form which is the simplest case of aggregation. To quantify the tendency to aggregate, we introduce the dimer debonding time and calculate the average magnetic force of attraction which results from correlations in the fluctuating moments. Neither quantity has any dependence on the magnetocrystalline anisotropy, which determines the rate of superparamagnetic reversals, and comparing with computed Néel relaxation times we show that this holds for both blocked and superparamagnetic particles. These results imply that the phenomenon of superparamagnetism does not affect aggregation. 
Because the key dimensionless parameter for the Néel relaxation of a lone SMNP and the one for magnetic attraction have the same size and temperature scaling, there is a strong correlation between superparamagnetism and colloidal stability, as observed experimentally, but no causal relation.} 

\end{tabular}

 \end{@twocolumnfalse} \vspace{0.6cm}

]

\renewcommand*\rmdefault{bch}\normalfont\upshape
\rmfamily
\section*{}
\vspace{-1cm}


\footnotetext{\textit{$^{a}$~DTU Physics, Technical University of Denmark, DK-2800 Kgs. Lyngby, Denmark.
E-mail: $\ddag$ fladu@dtu.dk, $\S$ fraca@fysik.dtu.dk}}
\footnotetext{\textit{$^{b}$~DTU Nanolab, Technical University of Denmark, 2800 Kgs. Lyngby, Denmark.}}
\footnotetext{\textit{$^{c}$~Department of Physics, University of Modena and Reggio Emilia, 41125 Modena, Italy.}}

\footnotetext{\dag~Electronic Supplementary Information (ESI) available: [details of any supplementary information available should be included here]. See DOI: 00.0000/00000000.}

\footnotetext{}

\section{Introduction}

Magnetic nanoparticles (MNPs) are of great scientific interest due to a range of applications in e.g.\ induction heating of chemical reactions\cite{roman_induction_2022,almind_optimized_2021,mortensen_direct_2017}, power electronics\cite{zambach_high-susceptibility_2023} and especially biomedicine\cite{pankhurst_applications_2003,pankhurst_progress_2009}, including magnetomechanical\cite{naud_cancer_2020} and hyperthermia\cite{thiesen_clinical_2008} treatment of cancer, targeted drug delivery\cite{beola_drug-loaded_2023}, novel immunoassays\cite{moyano_magnetic_2020}, magnetic particle imaging\cite{panagiotopoulos_magnetic_2015} and theranostic treatments\cite{coene_magnetic_2022}.

An important MNP subtype, used e.g.\ for biosensing and drug delivery\cite{xiao_superparamagnetic_2020,ha_recent_2018}, is the superparamagnetic nanoparticle (SMNP), i.e.\ a uniformly magnetised particle where the magnetisation does spontaneous reversals due to thermal fluctuations, hence the magnetic moment averages to zero over time. Technically all single-domain MNPs may experience thermal reversal events, however if the average reversal time greatly exceeds the measurement time the MNP is said to be blocked, rather than superparamagnetic.




In the synthesis and application of MNPs and other colloids, a key point is colloidal stabilisation, i.e.\ preventing spontaneous aggregation when the particles are in liquid suspension. This is especially challenging for magnetic colloids because of long-range magnetostatic attraction\cite{laurent_magnetic_2008}. It is generally accepted that SMNP suspensions are easier to stabilise than those with blocked particles. 
Typically the explanation cited is that since SMNP moments fluctuate randomly and time-average to zero, the magnetic attraction is largely absent\cite{xu_new_2013,olsvik_magnetic_1994,ha_recent_2018,neuberger_superparamagnetic_2005,nakata_chains_2008,kralj_magnetic_2015,gavilan_magnetic_2021}. However we find that this argument is at best inaccurate.

In the present study, we elucidate the fundamental link between superparamagnetism and aggregation, or rather absence thereof, by a series of Langevin dynamics simulations and statistical physics arguments. Specifically, we simulate MNP pairs and the dimer clusters they form, as this is the simplest, most fundamental aggregation process. 
To quantify the strength and stability of magnetic bonding, we use the dimer debonding time, i.e.\ how quickly the clusters break apart due to thermal fluctuations.
We prove that because of the magnetic dipolar interaction, nearby SMNP moments are correlated, so that even when thermal fluctuations dominate, there is a statistical tendency towards alignment and attraction. 
Moreover, our results show that the average force of attraction between colloidal MNPs and the simulated debonding times are both independent of anisotropy, and since raising anisotropy interpolates from superparamagnetic- to blocked particles, this implies that superparmagnetism does not affect aggregation. 

\section{Model \label{sec:model}}

\begin{figure}
    \centering
    \includegraphics[width=\columnwidth]{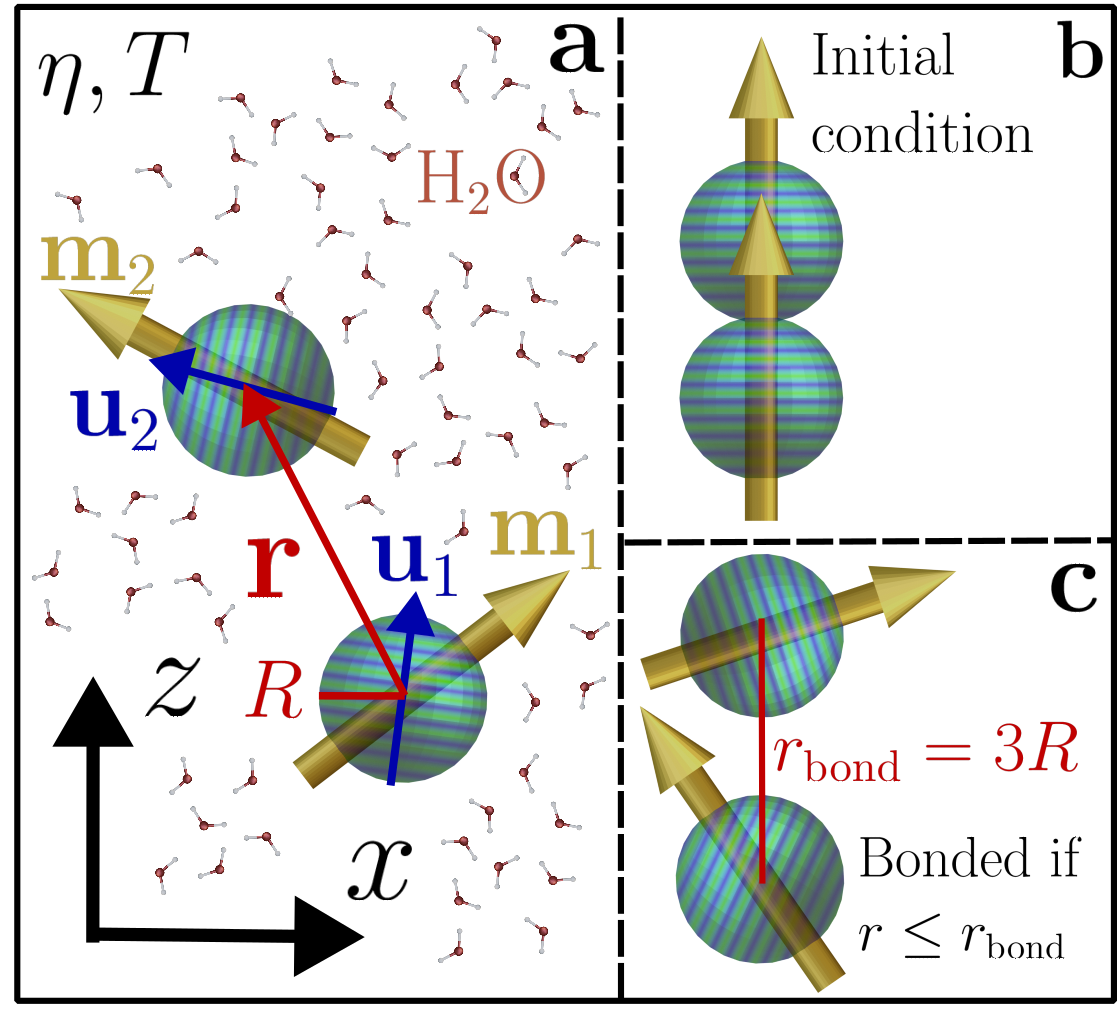}
    \caption{\textbf{a)} Illustration of system of interest, i.e.\ 2 identical magnetic nanoparticles (MNPs) in liquid suspension (illustrated with water). \textbf{a)} MNP radius $R$, relative displacement vector $\vb{r}$, normalised magnetic moments $\vb{m}_i=\bm{\mu}_i/MV$ and orientation vectors $\vb{u}_i$, which are given by the uniaxial anisotropy and is perpendicular to the blue/green stripes. \textbf{b)} The lowest energy state, i.e.\ particles in surface contact ($r=2R$) while both moments and anisotropy axes are aligned with the displacement vector ($\vb{m}_1=\vb{m}_2=\vb{u}_1=\vb{u}_2 =\vb{\hat{r}}=\vb{\hat{z}}$). This is the initial condition of all simulations except in \cref{fig:force_comparison_stat_phys}b where $r$ is varied. \textbf{c)} Bonding criterion.}
    \label{fig:Dimer_system_sketch}
\end{figure}

Depending on the details of solvent and particle surface, a number of non-magnetic effects may be critical in nanoparticle aggregation\cite{bishop_nanoscale_2009,israelachvili_intermolecular_2011,min_role_2008,russel_colloidal_1989}. Indeed hydrodynamic-\cite{satoh_brownian_1999,satoh_stokesian_1998}, electrostatic-\cite{chuan_lim_agglomeration_2012}, van der Waals (vdW)-\cite{durhuus_simulated_2021} and surface polymer\cite{novikau_influence_2020,rozhkov_self-assembly_2018,mostarac_characterisation_2020} interactions have all been applied when simulating aggregation in colloidal MNP suspensions. 

In the present study we seek to isolate the effect of thermal fluctuations on the magnetic attraction, in particular the role of superparamagnetism, so for clarity we consider a minimal model with only magnetic interactions and steric repulsion to limit particle overlap. Even then, simulating many-particle suspensions is computationally demanding, especially when including magnetic dynamics as this necessitates several orders of magnitude shorter timesteps\cite{berkov_langevin_2006,durhuus_conservation_2024}. 
Instead we exclusively simulate pairs of uniformly magnetised MNPs with uniaxial anisotropy,
and the dimer clusters they form, which also simplifies data interpretation.

For nanoparticles in liquid, the Reynolds number is generally low enough to neglect inertia\cite{purcell_life_1977}, 
which is equivalent to assuming zero mass density.
The energy for a pair of uniformly magnetised MNPs is then given by\cite{durhuus_conservation_2024}
\begin{align}
    E = E_\text{dip} + \sum_{i=1,2} E^\text{ani}_i  \label{eq:system_energy}
\end{align}
where the contribution from uniaxial anisotropy is
\begin{align}
    E^\text{ani}_i = -KV (\vb{u}_i \vdot \vb{m}_i)^2  \label{eq:E_ani}
\end{align} 
and the dipole-dipole interaction yields
\begin{align}
    E_\text{dip} = \frac{\mu_0 \mu^2}{4\pi r^3} [\vb{m}_1 \vdot \vb{m}_2 - 3 (\vb{m}_1 \vdot \vb{\hat{r}}) (\vb{m}_2 \vdot \vb{\hat{r}})],    \label{eq:E_dip}
\end{align}
with $\mu_0$ the vacuum permeability, $K$ the anisotropy constant, $V$ particle volume, $\mu = MV$ particle magnetic moment, $\vb{m}_i = \boldsymbol{\mu}_i/\mu$ the $i$'th normalised moment, $\vb{u}_i$ a unit vector parallel with the uniaxial anisotropy axis, $\vb{r}$ the displacement vector from MNP 1 to 2, $r = \abs{\vb{r}}$ the center-to-center distance and $\vb{\hat{r}} = \vb{r}/r$ the normalised displacement vector. 
Note that $E^\text{ani}_i$ is a symmetric, double-well potential where it is favourable for $\vb{m}_i$ to be fully parallel- or antiparallel with $\vb{u}_i$. See \cref{fig:Dimer_system_sketch} for an illustration.

From the system energy, \cref{eq:system_energy}, and well-established results on the drag and thermal fluctuations in liquid suspension, one can derive the full equations of motion\cite{durhuus_conservation_2024}. For the single-domain moments
\begin{align}
    \vb{\dot{m}}_i = -\gamma' \vb{m}_i \cross \vb{B}^\text{eff}_i - \alpha \gamma' \vb{m}_i \cross [\vb{m}_i \cross \vb{B}^\text{eff}_i]    \label{eq:LLG}
\end{align}
where $\alpha$ is the Gilbert damping parameter\cite{gilbert_phenomenological_2004}, $\gamma' = \gamma/(1 + \alpha^2)$ and the effective field is
\begin{align}
    \vb{B}^\text{eff}_i = \vb{B}^\text{dip}_i + \vb{B}^\text{ani}_i + \vb{B}^\text{th}_i + \alpha \gamma^{-1} \vb{m}_i \cross \bm{\omega}_i,
\end{align}
with contributions from dipolar interaction $\vb{B}^\text{dip}$, anisotropy $\vb{B}^\text{ani}$, thermal fluctuations $\vb{B}_i^\text{th}$ and the last term encodes the Barnett effect. The mechanical equations of motion are
\begin{align}
    \zeta_\text{r} \vb{\dot{u}}_i = \left[\gamma^{-1} \vb{\bm{\dot{\mu}}}_i + \bm{\mu} \cross \vb{B}_i^\text{dip} + \bm{\tau}^\text{th}_i\right] \cross \vb{u}_i,   \label{eq:rotation}
\end{align}
for rotation, and
\begin{align}
    \zeta_\text{t} \vb{\dot{r}}_i = \vb{F}^\text{dip}_i + \vb{F}^\text{WCA}_i + \vb{F}_i^\text{th}  \label{eq:translation}
\end{align}
for translation, where 
\begin{align}
    \zeta_\text{t} = 6\pi \eta R \qq{,} \zeta_\text{r} = 8\pi \eta R^3
\end{align}
are the friction coefficients for a sphere at low Reynolds number\cite{rubinow_transverse_1961}, $R$ is single particle radius, $\eta$ is dynamic viscosity, $\bm{\tau}^\text{th}$ and $\vb{F}^\text{th}$ are the thermal torque and force,
\begin{align}
    \vb{F}_\text{dip} &= \frac{3\mu_0 \mu^2}{4\pi} \frac{1}{r^4}[(\vb{m}_1 \vdot \vb{\hat{r}}) \vb{m}_2 + (\vb{m}_2 \vdot \vb{\hat{r}}) \vb{m}_1 
    \notag \\
    &\quad + (\vb{m}_1 \vdot \vb{m}_2)\vb{\hat{r}} - 5(\vb{m}_1\vdot \vb{\hat{r}})(\vb{m}_2 \vdot \vb{\hat{r}}) \vb{\hat{r}}]  \label{eq:Fdip_12}
\end{align}
is the dipole force from MNP 1 on MNP 2 and 
\begin{align}
    \vb{F}_\text{WCA} = \begin{cases}
    12 \epsilon_\text{WCA} \left[\frac{(2R)^{12}}{r^{13}} -  \frac{(2R)^6}{r^7}\right] 
    & r \leq 2R
    \\
    0 & r > 2R
    \end{cases} 
    \label{eq:F_WCA}
\end{align}
is the Weeks-Chandler-Anderson force from MNP 1 on MNP 2. $\vb{F}_\text{WCA}$ is a purely-repulsive interaction which kicks in when $r= 2R$, i.e.\ when the two particles are in surface contact, and prevents significant MNP overlap. We refer to Ref. \cite{durhuus_conservation_2024} for a full derivation and explicit formulas for the effective field contributions. Note that in neglecting inertia, we are using the overdamped limit. The Barnett effect is also negligible, but computationally cheap so we include it in simulations.

As for the thermal fluctuations, each vector component is a Gaussian distributed random variable with zero mean and no correlation between components. For both $\vb{F}^\text{th},\bm{\tau}^\text{th}$ and $\vb{B}^\text{th}$ the variance of each Cartesian component, $Q$, has the form 
\begin{align*}
\expval{Q(t) Q(t')} = 2k_B T C_Q \delta(t-t')
\end{align*}
where $\expval{...} $ is a canonical ensemble average, $k_B$ is Boltzmanns' constant, $T$ is temperature, the Dirac delta function $\delta(t-t')$ indicates no correlation between fluctuations at one point in time and another (zero autocorrelation), and the coefficients are
\begin{align}
    C_B = \frac{\alpha}{\gamma \mu} \qq{,} C_F = \zeta_\text{t} \qq{and} C_\tau = \zeta_\text{r}.
\end{align}
for the thermal field, force and torque respectively.


\subsection{Parameters}

The characteristic energy scales of the system are contained in the two dimensionless parameters
\begin{align}
    \sigma = \frac{KV}{k_BT} \qq{and} \lambda = \frac{\mu_0 \mu^2}{4\pi r^3 k_BT},  \label{eq:dimensionless_parameters}
\end{align}
where $\sigma$ is the ratio of anisotropy energy to thermal and $\lambda$ the characteristic dipole interaction energy at distance $r$ relative to thermal. Often we are interested in $\lambda$ for two MNPs in surface contact:
\begin{align}
    \eval{\lambda}_{r=2R} = \frac{\pi \mu_0 M^2 R^3}{18 k_BT}. \label{eq:lambda_surface_contact}
\end{align}

We simulate 4 different values of $K$ to vary $\sigma$ and when changing $\lambda$ we vary either $R, T$ or $M$ with the other two kept at default values (see \cref{tab:parameters}). The default parameters correspond to typical 10 nm diameter, iron-oxide particles at ambient temperature and $\eta = 1\:\mathrm{mPa\cdot s}$ matches water or oleic acid under ambient conditions.

\def\arraystretch{1.2}%
\begin{table}[tbh]
    \centering
    \begin{tabular}{|c|c|c|c|c|}
        \hline
        Symbol & Description & Unit & Default & Range  \\
        \hline
        $R$ & Radius & $\mathrm{nm}$ & $5$ & $1..10$ \\
        \hline
        $T$ & Temperature & $\mathrm{K}$ & $300$ & $30..1500$ \\
        \hline
        $M$ & Magnetisation & $\mathrm{kA / m}$ & $400$ & $200..1200$ \\
        \hline
        $K$ & Anisotropy constant & $\mathrm{kJ / m^{3}}$ & $20$ & $0, 10, 20, 40$ \\
        \hline
        $\sigma$ & $KV/k_B T$ & $1$ & $2.53$ & $0..20.2$ \\
        \hline
        $\lambda$ & $\pi \mu_0 M^2 R^3 / (18k_B T)$
        & 1 & $1.06$ & 0.01..10.6 \\
        \hline
        $\eta$ & Dynamic viscosity & $\mathrm{mPa\cdot s}$ & $1$ & $1$ \\
        \hline
        $\epsilon_\text{WCA}$ & WCA parameter & $\mathrm{J}$ & $10^{-17}$ & $10^{-17}$ \\
        \hline
        $\alpha$ & Gilbert damping & $1$ & $0.01$ & $0.01$ \\
        \hline
        $\Delta t$ & Timestep & $\mathrm{ps}$ & 10 & 10 \\
        \hline
    \end{tabular}
    \caption{Simulated parameters. Unless otherwise noted we vary $\sigma$ via $K$ and $\lambda$ via either $R, T$ or $M$ while keeping the other two at their default value. The default parameters are representative of iron oxides and the ranges ensure $\sigma, \lambda$ vary from $\ll 1$ to $\gg 1$.}
    \label{tab:parameters}
\end{table}

\section{Superparamagnetic relaxation time of dimers \label{sec:superparamagnetism_of_dimers}}

\begin{figure*}[!ht]
    \centering
    \includegraphics[width=\textwidth]{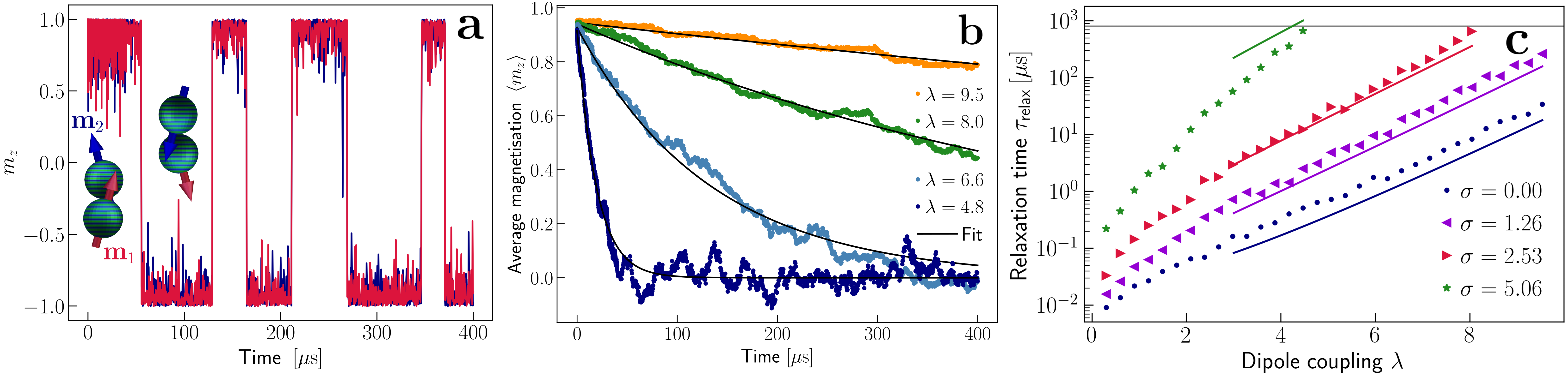}
    \caption{\textbf{(a)} Example simulation of mechanically fixed dimer with $M=900\:\mathrm{kA/m}\: (\lambda = 5.36)$ and $K=20\:\mathrm{kJ/m^3}\: (\sigma = 2.53)$, showing $m_z(t)$ for both constituent MNPs (red and blue data).
    \textbf{b)} Average magnetisation used to determine $\tau_\text{relax}$. Each colored curve is the average of 250 dimers with identical parameters; here $K=20\:\mathrm{kJ/m^3}$ and $M$ is varied between curves. Black lines are exponential fits to \cref{eq:tau_relax}.
    \textbf{c)} Superparamagnetic relaxation time for the dimer as a function of the anisotropy and dipole parameters. $\lambda$ is varied via $M$. Each symbol corresponds to a 250-dimer fit and the solid lines are the predictions of the $\lambda \gg 1$ theory developed in \cref{appsec:relaxation_time_in_strong_coupling_limit}, shown for $\lambda \geq 3$. The gray line indicates a cutoff at $800\: \mathrm{\mu s}$, i.e.\ twice the simulation time, above which a longer simulation is required to determine $\tau_\text{relax}$ accurately.
    }
    \label{fig:tauRelax}
\end{figure*}

When an ensemble of MNPs is perturbed, e.g.\ by an external magnetic field, the magnetic moments will decay exponentially back to thermal equilibrium by Néel relaxation in a solid substrate\cite{neel_theorie_1949,brown_thermal_1963} (magnetic dynamics), or combined Néel and Brownian relaxation in fluid suspension\cite{shliomis_frequency_1993,stepanov_combined_1991,kroger_combined_2022} (magnetic and mechanical rotation). Here we use the characteristic Néel relaxation time, $\tau_\text{relax}$ to quantify the degree of superparamagnetism in a dimer, including the effect of magnetostatic interactions.

There exists a large body of analytical and numerical work on $\tau_\text{relax}$ for a lone MNP, comprehensively reviewed in Ref. \cite{coffey_thermal_2012}. We note in particular that Brown derived a neat closed-form solution in the limit of a high uniaxial anisotropy barrier\cite{brown_thermal_1963}, Coffey et.\ al. derived a more cumbersome general solution\cite{coffey_exact_1994} and Eisenstein and Aharoni solved the high-barrier limit for cubic anisotropy\cite{eisenstein_asymptotic_1977}. 
For MNP aggregates, there are a number of more recent numerical studies on magnetic relaxation in small compact clusters\cite{ilg_equilibrium_2017,hovorka_role_2014,hovorka_thermal_2017} and particle chains\cite{hovorka_role_2014,anand_relaxation_2019,leliaert_regarding_2014,anand_controlling_2022}, however we are not aware of any exact, analytical results for the dimer aggregate, so we extract $\tau_\text{relax}$ from simulations.

We consider a dimer which is mechanically fixed in the initial, fully aligned state (\cref{fig:Dimer_system_sketch}b), i.e.\ $\vb{u}_1=\vb{u}_2=\vb{\hat{r}}=\vb{\hat{z}}$ and $r=2R$ at all times. The only equation of motion is then the LLG, \cref{eq:LLG}. We find that in this setting, the method of Chalifour et.\ al.\cite{chalifour_magnetic_2021} to compute relaxation times for single particles readily generalises to dimers.

The idea is to simulate an ensemble of identical, isolated dimers, here 250.
As time progresses some magnetic moments will spontaneously flip ca.\ $180^\circ$ to the opposite anisotropy energy minimum, which is known as interwell motion. Then the variance in the distribution of moment directions grows until there is an equal number of up/down moments and the ensemble average $\expval{\vb{m}}$ is zero. Because the reversals on different dimers are independent, discrete, random events, they obey the Poisson distribution. Naturally there are correlations between the two moments within a dimer, but in a large ensemble this averages out.
We prove in \cref{appsec:Néel_relaxation_statistics} that Poisson distributed reversals entails an exponential decay of $\expval{m_z}$ in time:
\begin{align}
    \expval{m_z}(t) = \ee^{-t/\tau_\text{relax}}   \label{eq:tau_relax}
\end{align}
which defines $\tau_\text{relax}$. 

As noted in Ref. \cite{chalifour_magnetic_2021} there is a small, initial drop in $\expval{m_z}$ on a ns timescale due to intrawell motion (small fluctuations within the groundstate energy well). This intrawell relaxation is consistent with the low-temperature magnetisation model of Mørup et.\ al.\ \cite{morup_uniform_2010, morup_experimental_2007}, and should be filtered out since \cref{eq:tau_relax} only models interwell reversals. However in the low barrier limit ($\sigma, \lambda \ll 1$) intra- and interwell motion are indistinguishable. Therefore, when \cref{eq:tau_relax} yields $\tau_\text{relax} \leq 100\:\mathrm{ns}$ we fit the full curve while for $\tau_\text{relax} > 100\:\mathrm{ns}$ we only use the $t \geq 10 \:\mathrm{ns}$ part, i.e.\ we redo the fit using
\begin{align*}
    \expval{m_z}(t) = \expval{m_z}(t_0)\ee^{-(t-t_0)/\tau_\text{relax}},
\end{align*}
where $t_0 = 10\:\mathrm{ns}$. 
To explain the value of $t_0$, we note that without anisotropy or interactions, the characteristic timescale for magnetic relaxation is\cite{berkov_langevin_2006} $\mu/(\alpha \gamma k_B T)$, which is $86\:\mathrm{ns}$ at the highest magnetisation (slowest relaxation), however intrawell relaxation constitutes a few $\%$ decrease in $\expval{m_z}$ rather than full moment randomisation, so a small fraction of the time is required.

The method is exemplified in \cref{fig:tauRelax}a-b. \Cref{fig:tauRelax}a shows how the two moments on a dimer suddenly flip at random intervals, and due to a relatively strong dipolar coupling of $\lambda = 5.4$ in this example the flips are synchronised, indicating strongly correlated fluctuations. 
Note that while the components along the interparticle axis are aligned, the transverse components tend to anti-align, in agreement with \cref{subsec:correlation_measures_and_force_of_attraction}. 
From \cref{fig:tauRelax}b we see that the ensemble-averaged moment, $\expval{m_z}(t)$, is well-described by exponential fits when starting the fit at $t=10\:\mathrm{ns}$, even when the relaxation time exceeds the simulation time. Moreover a stronger dipole coupling increases $\tau_\text{relax}$, and it also appears to decrease the amplitude and frequency of fluctuations in $\expval{m_z}$.

In \cref{fig:tauRelax}c we show $\tau_\text{relax}$ as a function of $\sigma$ and $\lambda$. The linear growth on a single-logarithmic scale indicates that $\tau_\text{relax}$ is exponentially dependent on $\lambda$. 
In \cref{appsec:relaxation_time_in_strong_coupling_limit} we derive a model of the relaxation time in the strong coupling limit $\lambda \gg 1$:
\begin{align}
    \tau_\text{relax} = \frac{\sqrt{\pi}\mu}{4 \alpha \gamma' k_BT} \frac{\ee^{\Tilde{\sigma}}}{\Tilde{\sigma}^{3/2}} \qq{where} \Tilde{\sigma} = 2\sigma + \lambda. \label{eq:tau_relax_optimal_path}
\end{align}
The derivation assumes that due to a strong dipole coupling the two moments are perfectly correlated, so $\vb{m}_2$ is effectively a function of $\vb{m}_1$ and we get the same degrees of freedom as the single particle case. Indeed the end result and most of the argument is identical to that for a single particle with uniaxial anisotropy\cite{brown_thermal_1963,coffey_thermal_2012} except that $\sigma \xrightarrow{} 2\sigma + \lambda$. The reason is that when the moments flip together, as in \cref{fig:tauRelax}a, thermal fluctuations have to overcome the dipole coupling and both anisotropy barriers simultaneously. 
Because of this doubling of the anisotropy barrier, high $\sigma$ favours individual reversals, which is why the model fails so spectacularly at high $\sigma$ and low $\lambda$ (see the green curve of \cref{fig:tauRelax}c).
It also undershoots a bit at low $\sigma$, however for a model with no fitting parameters there is good agreement with simulations in the appropriate limit.

In summary, the relaxation time has an Arrhenius-style exponential dependence on both anisotropy and dipolar energy parameters, and in the regime dominated by dipole coupling ($\lambda \gg 1, \sigma$), the tendency towards synchronous reversal increases the effective anisotropy barrier.

One can also describe superparamagnetism in terms of the reversal time $\tau_\text{rev}$, defined as the average time between moment reversals. Counting the time between reversals directly, or equivalently the average time before the first reversal as in Ref.\ \cite{kalmykov_damping_2010}, measures $\tau_\text{rev}$. It happens that for a Poisson distribution $\tau_\text{rev} = 2\tau_\text{relax}$ (see \cref{appsec:Néel_relaxation_statistics}) so with an appropriate factor of $1/2$ the two timescales coincide, which is why Ref. \cite{kalmykov_damping_2010} and Ref. \cite{chalifour_magnetic_2021} both agree with the analytical relaxation time from Ref. \cite{brown_thermal_1963}. An important difference is that the notion of discrete reversal events is only well-defined in the high barrier case, so computing $\tau_\text{rev}$ requires $\sigma$ or $\lambda$ to be $\gg 1$, while $\tau_\text{relax}$ is meaningful at all $\sigma,\lambda$ values. Also $\tau_\text{relax}$ includes relaxation from intrawell motion while $\tau_\text{rev}$ does not.

For a measurement time of $\tau_m$, the relaxation time is related to the experimentally relevant blocking temperature, $T_B$, by $\tau_m = a \tau_\text{relax}(T_B)$ where $a$ is a constant (typically 100) \cite{blundell_magnetism_2001}. For \cref{eq:tau_relax_optimal_path} this amounts to
\begin{align*}
    T_B = \frac{1}{k_B \ln\left(\frac{\tau_m}{a \tau_0}\right)}\left[2KV + \frac{\mu_0 \mu^2}{4\pi r^3} \right] \qq{where} \tau_0 = \frac{\sqrt{\pi} \mu}{4\alpha \gamma' k_B T_B \Tilde{\sigma}^{3/2}},
\end{align*}
which is consistent with Ref. \cite{rivas_rojas_comparison_2022}.

\section{Dimer debonding time \label{sec:dimer_bonding_time}}

To study the strength and stability of the dimer aggregates, which are only bonded by magnetostatic attraction, we simulate how long it takes on average for the bond to break, in the sense that the particle distance $r$ exceeds a critical bonding distance $r_\text{bond}$ (cf. \cref{fig:Dimer_system_sketch}c). We refer to this as the debonding time, $\tau_\text{debond}$. We only consider the first breaking of each bond, not recombination events or repeated breaking. In the study of stochastic systems, this is referred to as a first-passage problem\cite{redner_guide_2001,bray_persistence_2013}.

We consider an ensemble of MNP dimers initiated in the fully aligned, lowest energy state as in \cref{fig:Dimer_system_sketch}b, simulating the time evolution of each MNP pair, now with full magnetic and mechanical degrees of freedom. 
The dipole potential goes asymptotically to 0, so the bond criterion is ambiguous. We chose $r_\text{bond} = 3R$.

For the choice of the WCA parameter $\epsilon_\text{WCA}$, we would ideally use a hard-sphere potential, i.e.\ set $\epsilon_\text{WCA} = \infty$, so the equilibrium distance is exactly $r=2R$, however this is numerically unstable. Conversely if the potential is too soft (low $\epsilon_\text{WCA}$) there will be significant particle overlap, decreasing $r$ and increasing $\lambda \sim 1/r^3$. In the supplementary information\cite{supplementary_information} we show that $\epsilon_\text{WCA} = 10^{-17}\:\mathrm{J}$ is high enough to reach convergence.

Let $N_u(t)$ be the number of undivided pairs, i.e.\ bonds that have not been broken at any point. Bond breaks are independent, discrete, random events, so analogously to radioactive decay or chemical bonds, the rate of bond breaking in MNP dimers is proportional to the number of remaining bonds:
\begin{align}
    \dv{t}N_u(t) = -\frac{1}{\tau_\text{debond}} N_u(t) \Rightarrow N_u(t) = N_0 \ee^{-t/\tau_\text{debond}},   \label{eq:diff_N_u}
\end{align}
where $N_0$ is the number of MNP pairs. We denote the characteristic time $\tau_\text{debond}$ as the magnetic debonding time. Analogously to the superparamagnetic relaxation time we determine $\tau_\text{debond}$ by an exponential fit to $N_u(t)/N_0$ where $N_0 = 250$.

\begin{figure}
    \centering
    \includegraphics[width=0.95\columnwidth]{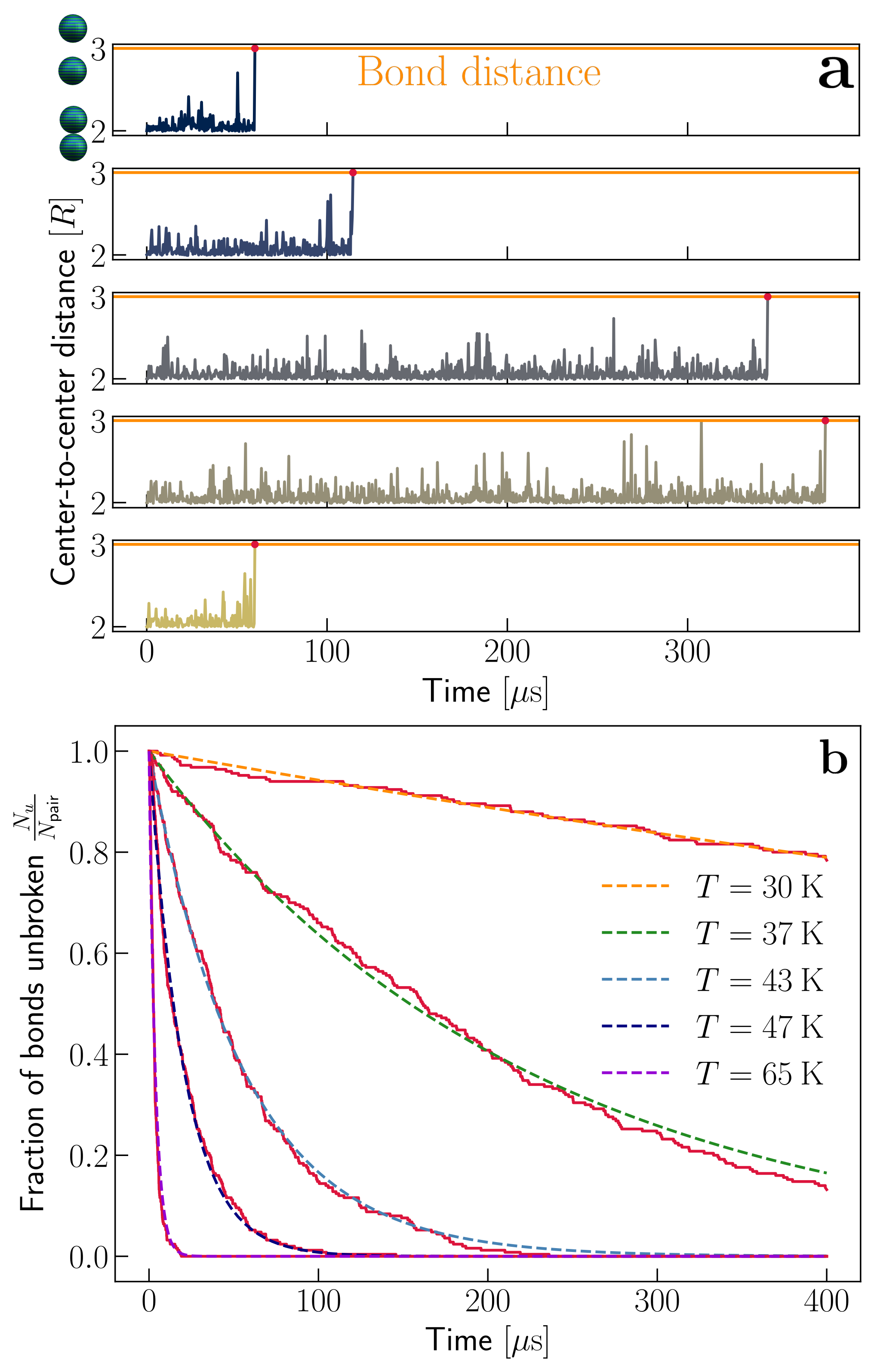}
    \caption{\textbf{a)} Simulations of the center-to-center distance $r$ vs.\ time for 5 MNP pairs. Orange lines indicate the critical bond distance and red dots the moments of debonding, $t_\text{break}$. 
    Initial- and debonding distances are illustrated in the top left. 
    \textbf{b)} Data (solid lines) with exponential fits (dashed lines) for $K = 20\:\mathrm{kJ/m^3}\: (\sigma=2.53)$ and $\lambda$ varied via temperature. 
    Looking closely the data (solid lines) change in discrete steps as individual pairs debond.}
    \label{fig:fitting_method}
\end{figure}

The method is illustrated in \cref{fig:fitting_method}, with single-dimer simulations in \cref{fig:fitting_method}a and ensemble averaged time-evolution in \cref{fig:fitting_method}. From \cref{fig:fitting_method}b we see extremely clear exponential fits, even for $T=30\:\mathrm{K}$ which corresponds to the longest debonding time in our simulations. In the supplementary information\cite{supplementary_information} we present several convergence tests and data with different $r_\text{bond}$ values to further validate the method. 

We evaluate pair distances as part of the core computational loop; saving the time-coordinates $t_\text{break}$ where each dimer first exceeds $r_\text{bond}$. Thus we get $t_\text{break}$ with the same $0.01 \: \mathrm{ns}$ resolution as the timestep. If $\tau_\text{debond}$ is instead computed as a post processing step, and an MNP pair breaks apart and recombines between two saved datapoints, the break is missed, which skews the counting statistics and introduces an artificial $\lambda$ dependence.

\begin{figure*}
    \begin{minipage}{0.5\textwidth}
        \includegraphics[width=\textwidth]{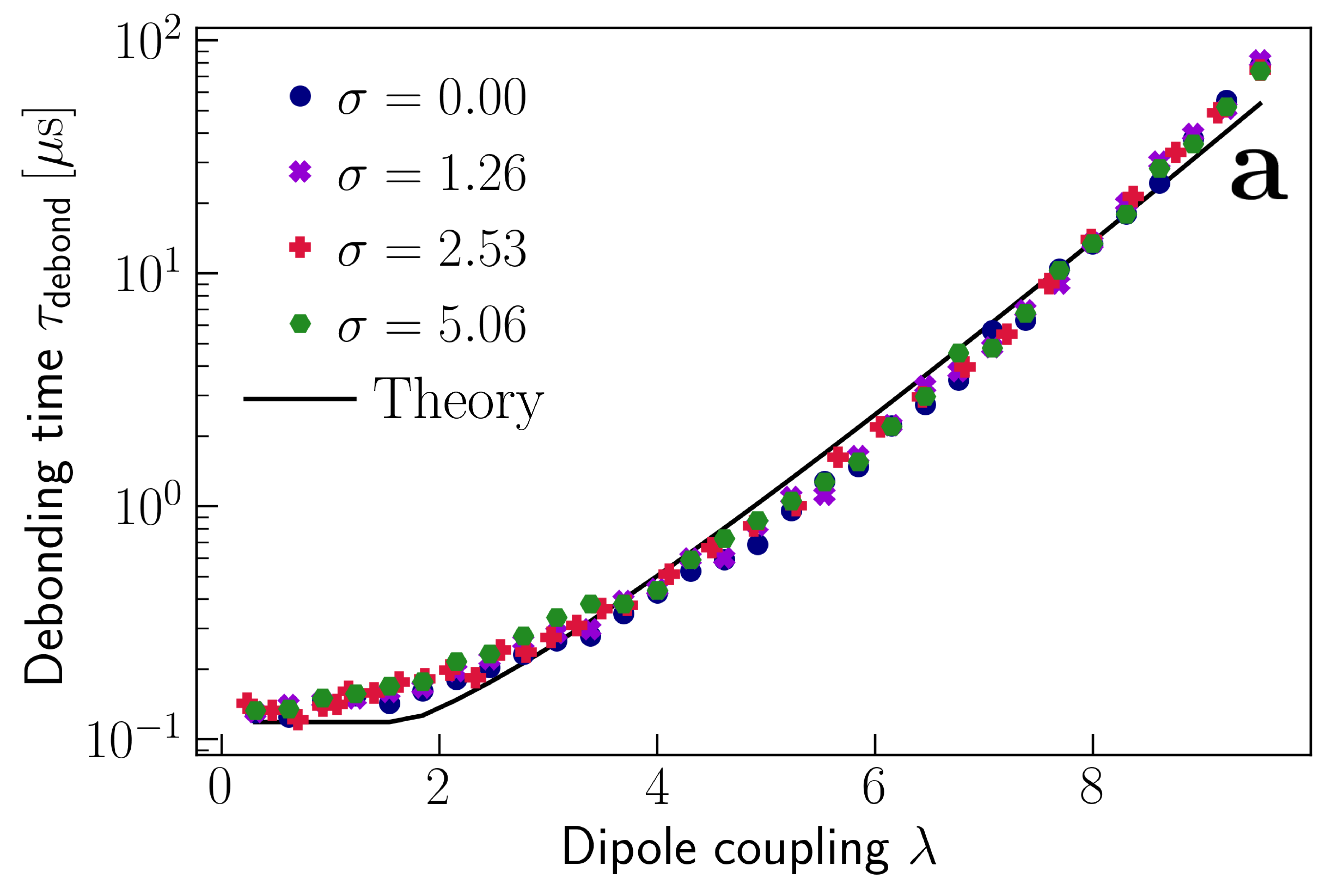}
    \end{minipage}%
    \begin{minipage}{0.5\textwidth}
        \includegraphics[width=\textwidth]{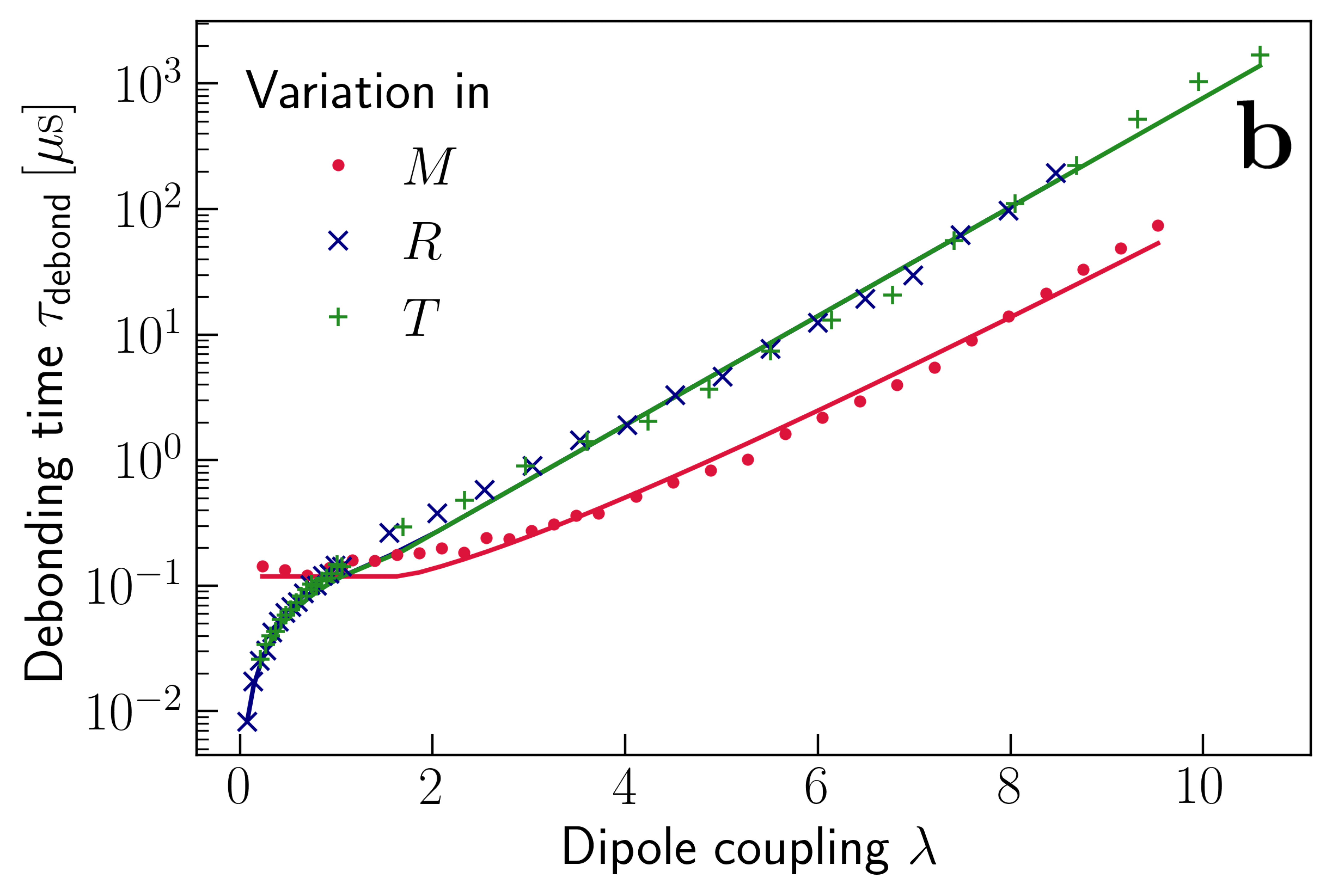}
    \end{minipage}
    \centering
    \includegraphics[width=\textwidth]{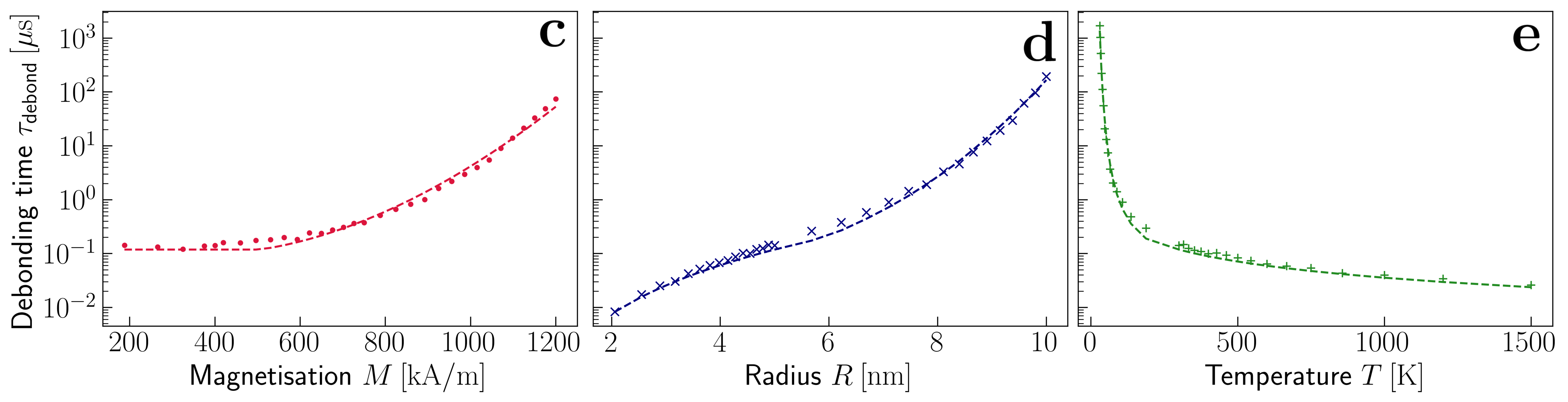}
    \caption{Average bond lifetime $\tau_\text{debond}$ vs. relevant parameters. Each symbol is a datapoint from fitting the debonding of 250 dimers (cf. \cref{fig:fitting_method}). Solid lines are the theoretical model of \cref{eq:bond_time_expression} where we used $(A,B) = (2.5, 7)$ as fitting parameters throughout. In \textbf{a)} $\lambda$ is varied via $M$. In \textbf{b)} the colours correspond to whether $\lambda$ was varied via $M, R$ or $T$. \textbf{c-e)} shows the same data as in \textbf{b)}. Parameter ranges and default values are given in \cref{tab:parameters}.}
    \label{fig:lifetime_simulations}
\end{figure*}

In \cref{fig:lifetime_simulations} the debonding time $\tau_\text{debond}$ is shown as a function of $M, T, R$ and $K$. From \cref{fig:lifetime_simulations}a we see a clear exponential dependence on the dipole coupling strength $\lambda$, but $\sigma$ and by extension anisotropy and superparamagnetism, have no influence on debonding time. At first glance the apparent irrelevance of $\sigma$ might be a matter of timescales, however comparing with \cref{fig:tauRelax}c, we see that the $\sigma = 0$ simulations generally have $\tau_\text{debond} \sim \tau_\text{relax}$ while for $\sigma=5.06$ we have $\tau_\text{debond} \ll \tau_\text{relax}$. 
Thus if superparamagnetic reversals had any impact on bond strength and stability it should appear in \cref{fig:lifetime_simulations}a, yet the effect is absent. We discuss the statistical physics underpinning this surprising result in \cref{sec:Equilibirum_statistical_physics}. 

Regarding the $\lambda$ dependence, when a bond breaks, it is because the system crosses an energy barrier (here the dipole potential) by means of thermal fluctuations, which is the same basis as the Arrhenius law of reaction kinetics\cite{gibbs_sufficient_1972,arrhenius_uber_1889}. Such exponential dependence on the ratio of barrier height to thermal energy is also known from first passage problems in general\cite{redner_guide_2001,bray_persistence_2013} when the system is escaping out of a potential well.

As for the prefactor, the only relevant parameters are $M, R, T$ and $\eta$. The only combinations of these parameters with units of time are
\begin{align}
    \tau_\text{Brown} = \frac{\eta R^3}{k_B T} \qq{and} \tau_\text{dip} = \frac{\eta}{\mu_0 M^2},
\end{align}
so by dimensional analysis these are the only characteristic timescales possible.
If we disregard the magnetisation, $\tau_\text{Brown}$ is the sole time constant, so it must correspond to Brownian motion. We hypothesise that for small $\lambda$ the energy barrier is negligible and $\tau_\text{debond}$ is diffusion limited, while for high $\lambda$ the barrier escape time dominates. Similarly to how potential gradients determine the prefactor in a Kramers' escape problem, $\tau_\text{dip}$ must be related to the shape of the dipole potential. We propose the final debonding time expression
\begin{align}
    \tau_\text{debond} = \max(A \tau_\text{Brown},\: B \tau_\text{dip} \ee^{\lambda}), \label{eq:bond_time_expression}
\end{align}
where $A, B$ are dimensionless fitting parameters which depend on the chosen $r_\text{bond}$.
$(A, B) = (2.5, 7)$ work well for the simulated data when varying both $M, T$ and $R$. From \cref{fig:lifetime_simulations} we see that \cref{eq:bond_time_expression} captures the dependence of debonding time on the relevant parameters for both low and high $\lambda$ with moderate discrepancy in the intermediate regime of $\lambda \sim 2$. 
In the supplementary information\cite{supplementary_information} we also verify the linear dependence on $\eta$.

Referring to \cref{fig:lifetime_simulations}c-e, $R=5\:\mathrm{nm}$ iron oxide particles ($M\sim 400\:\mathrm{kA/m}$) at room temperature are diffusion limited, indicating essentially no magnetic bonding.
Lowering temperature below $300\:\mathrm{K}$ can massively increase debonding time while raising it has a comparatively minor impact. The exact magnetisation is unimportant in the diffusion limited regime, while particle size is significant at all length scales because both $\tau_\text{Brown}$ and $\eval{\lambda}_{r=2R}$ scale as $R^3$.

We note that the notion of debonding time readily generalises to non-magnetic colloids. With the present implementation, once two MNPs debond, Brownian motion may take them so far apart they essentially cease to interact. However with periodic or reflecting boundary conditions, it should be possible to treat the dissociation and recombination rates on an equal footing. Then one could compute reaction rates and equilibrium constants analogous to those of chemical kinetics, but for colloids.

\section{Equilibrium statistical physics \label{sec:Equilibirum_statistical_physics}}



\subsection{Irrelevance of anisotropy}

In the preceding sections we computed both Néel relaxation times and bond lifetimes, finding unequivocally that anisotropy, and hence superparamagnetism, is irrelevant for bond stability and consequently does not affect aggregation. Here we resort to equilibrium statistical physics to explain this result and infer its range of validity.

Because particle number is conserved, the statistics are fully described by the partition function $Z$, which is a sum or integral over all system states weighted by the Boltzmann factor\cite{blundell_concepts_2010} $\ee^{-E/k_BT}$. 
It follows from \cref{eq:system_energy} and translational symmety that the partition function for a 2-MNP system may be written
\begin{align}
    Z = \int \dd\vb{r} \dd\vb{u}_1 \dd\vb{u}_2 \dd\vb{m}_1 \dd\vb{m}_2 \ee^{-E/k_B T} = I_0^2 \int \ee^{-\lambda \Tilde{E}_\text{dip}} \dd \vb{r} \dd\vb{m}_1 \dd\vb{m}_2  \label{eq:partition_function}
\end{align}
where  
\begin{align}
I_0 =  \int \ee^{\sigma (\vb{u} \vdot \vb{m})^2} \dd \vb{u}, 
    \label{eq:partition_function_integral}
\end{align}
and we defined $\Tilde{E}_\text{dip} = E_\text{dip} / \lambda$.

Importantly $I_0$ is not a function of $\vb{m}$, since the integrand only depends on the relative orientation of $\vb{m}$ wrt.\ $\vb{u}$ and we integrate $\vb{u}$ over all directions. 
This is how the $\vb{u}$ integrals, which contain all the $\sigma$-dependence, factorise from the rest of $Z$.

The probability of a given configuration in thermal equilibrium is $P(\vb{u}_1, \vb{u}_2, \vb{m}_1, \vb{m}_2, \vb{r}) = \frac{1}{Z} \ee^{-E/k_BT}$. We can calculate the probability of just the moment directions and displacement by integrating over all mechanical orientations:
\begin{align}
    P(\vb{m}_1, \vb{m}_2, \vb{r}) &= \int P(\vb{u}_1, \vb{u}_2, \vb{m}_1, \vb{m}_2, \vb{r}) \dd \vb{u}_1 \dd \vb{u}_2
    \notag \\
    &= \frac{\ee^{-\lambda \Tilde{E}_\text{dip}}}{\int \ee^{-\lambda \Tilde{E}_\text{dip}} \dd \vb{r} \dd\vb{m}_1 \dd\vb{m}_2}.
\end{align}
We see that because of how $Z$ factorises, all dependence on $\sigma$, and by extension $K$, cancels out. In other words, the moment configuration and spatial distribution of MNPs in fluid suspension is independent of anisotropy. One consequence is that the particle anisotropy has no impact on the formation or stability of MNP aggregates. However when $\sigma \xrightarrow{} \infty$ the magnetic moment is locked to the anisotropy axis, while in the opposite limit of $\sigma = 0$ a lone particle is guaranteed to be superparamagnetic. In conclusion, whether or not MNPs are superparamagnetic does not affect their tendency to aggregate in fluid suspension.

The same argument applies for all higher order anisotropies and an arbitrary number of interacting MNPs, however we do require mechanical freedom of rotation for the $\vb{u}$-integral to factorise.

A useful interpretation is that when the MNPs are fixed in space, the sample magnetisation can only change by crossing the anisotropy energy barriers, however in a fluid the magnetic moments can follow the mechanical rotation, enabling the same degrees of freedom to relax 
into thermal equilibrium 
without crossing any energy barriers. The mechanical rotation is several orders of magnitude slower than magnetic dynamics\cite{durhuus_conservation_2024,berkov_langevin_2006}, however this is irrelevant for thermal equilibrium properties. Hence given enough time to relax, the system statistics are the same as if the anisotropy barrier was absent ($\sigma=0$), i.e.\ as if all the MNPs were perfectly soft magnets.


This statistical decoupling of mechanical rotation and the other degrees of freedom was also noted by Elfimova et.\ al.\cite{elfimova_static_2019} when studying the static magnetic susceptibility of SMNP suspensions. Their conclusion was likewise that $\sigma$ does not affect the susceptibility of ferrofluids, but does matter when the particles are immobilised in a solid. 

In susceptibility measurements, there is by necessity an applied field, hence even a sample of SMNPs will have a net average magnetisation. In zero-field cases however the magnetic moments of lone SMNPs time-average to zero, so how can aggregation occur? The answer is that when two SMNPs approach eachother, the dipole-dipole interaction induces correlations in the magnetic moment directions, which on average leads to magnetostatic attraction. 
Below, we explicitly evaluate these correlations and the resulting force.

\subsection{Correlation measures and force of attraction \label{subsec:correlation_measures_and_force_of_attraction}}

The average force of attraction between two MNPs at a given distance is the ensemble average of the dipole force, \cref{eq:Fdip_12}, at fixed distance, $\expval{F_\text{dip}}(r)$. In other words, the thermally weighted average over all combinations of $\vb{u}_1, \vb{u}_2, \vb{m}_1$ and $\vb{m}_2$ at constant $r$.
Referring to \cref{eq:Fdip_12} this amounts to averaging a number of dot products such as $\expval{\vb{m}_1 \vdot \vb{m}_2}$ and $\expval{(\vb{m}_1 \vdot \vb{\hat{r}})(\vb{m}_2 \vdot \vb{\hat{r}}_2)}$. The former measures the moment-moment correlation and the latter correlations along the interparticle axis. To these we add the correlation between moment and anisotropy axis $\sqrt{\expval{(\vb{u} \vdot \vb{m})^2}}$ (squaring is necessary because $\expval{\vb{m} \vdot \vb{u}} = 0$), and the transverse correlation $\expval{\vb{m}_{1\perp} \vdot \vb{m}_{2\perp}}$ where $\vb{m}_{i\perp}$ is the normalised component of $\vb{m}_i$ perpendicular to $\vb{r}$.

In \cref{appsec:correlation_measures_in_weak_coupling_limit,appsec:effective_dipole_force} we derive simple, analytical expressions for these correlation measures and the resulting $\expval{F_\text{dip}}$ in the weak coupling (low $\lambda$) regime. To second order in $\lambda$
\begin{align}
    \expval{\vb{F}_\text{dip}} = - \frac{\mu_0 \mu^2}{2\pi d^4} \lambda \vb{\hat{r}} = -\frac{\mu_0^2 \mu^4}{8\pi^2 r^7} \vb{\hat{r}} \label{eq:F_dip_avg} 
\end{align}
with the effective potential 
\begin{align}
U_\text{eff} = - \frac{\mu_0^2 \mu^4}{48\pi^2 r^6}. \label{eq:averaged_attraction}
\end{align}
We define the effective potential by $-\grad U_\text{eff} = \expval{\vb{F}_\text{dip}}$, because ensemble averaging the dipole potential yields $\expval{E_\text{dip}} = 2 U_\text{eff}$ so the average force is not the gradient of the average potential. 

We note that the force scales as $1/r^7$. At high $\lambda$, MNPs tend to align their moments before colliding \cite{durhuus_simulated_2021}, producing an $r^{-4}$ force dependence, so thermal fluctuations both reduce the effective force and make the attraction more localised. Also the $\mu^4$ factor indicates a very strong dependence on magnetisation and particle size. 
Because both the average force and the translational Brownian motion are independent of $\sigma$, anisotropy does not affect the timescale of aggregation or the resulting structures. Consequently the results of studies like \cite{durhuus_simulated_2021} which simulate MNP aggregation under the RDA approximation (moments locked to the anisotopy axis) are directly applicable to superparamagnetic MNPs.

\begin{figure}
    \centering
    \includegraphics[width=\columnwidth]{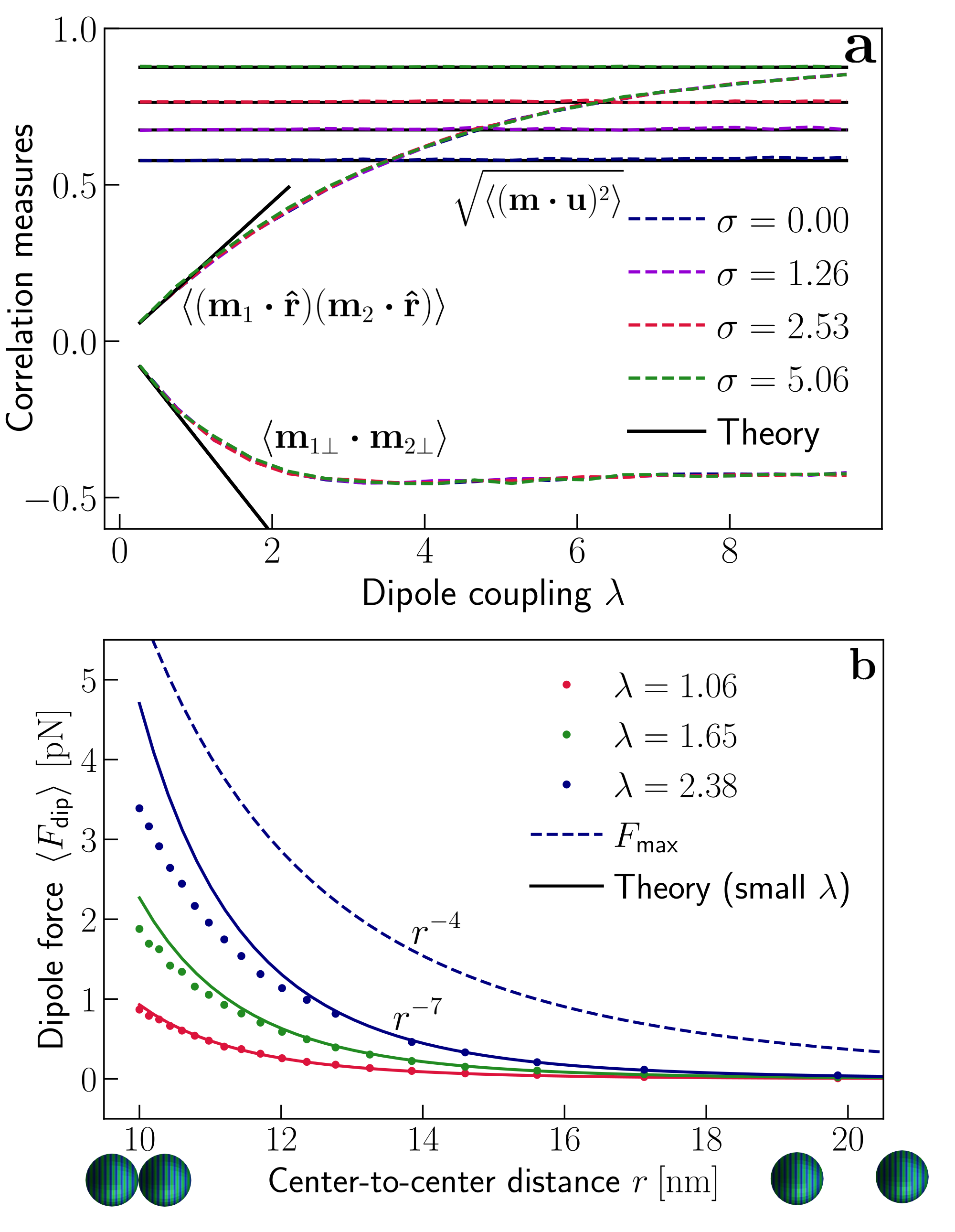}
    \caption{\textbf{a)} Correlation measures vs.\ $\lambda$ and $\sigma$. Dashed lines are simulation results with positions fixed in surface contact ($r=2R$). The black curves are theory results given in \cref{appsec:correlation_measures_in_weak_coupling_limit}, which for $\expval{(\vb{m}_1 \vdot \vb{\hat{r}})(\vb{m}_2 \vdot \vb{\hat{r}})}$ and $\expval{\vb{m}_{1\perp}\vdot \vb{m}_{2\perp}}$ are derived for low $\lambda$, hence the cutoff at $\lambda=2$. 
    \textbf{b)} Thermally averaged dipole force vs.\ distance. The given lambda values correspond to surface contact. Dots are numerical ensemble averages (cf. \cref{appsec:numerical_ensemble_average}), solid lines the low-$\lambda$ theory of \cref{eq:F_dip_avg} and the dashed line the force when $\vb{m}_1$ and $\vb{m}_2$ are perfectly aligned for $\lambda=2.38$. 
    The different $r$-dependencies for maximal- and thermally averaged force is indicated.}
    \label{fig:force_comparison_stat_phys}
\end{figure}

In \cref{fig:force_comparison_stat_phys} we compute correlation measures and the averaged force numerically, by time-averaging over simulations with fixed distance but both magnetic and mechanical rotation (see \cref{appsec:numerical_ensemble_average} for details). In \cref{fig:force_comparison_stat_phys}a we see that $\expval{\vb{m}_{1\perp} \vdot \vb{m}_{2\perp}} < 0$, so the transverse components tend to anti-align. This stems from the first term in $E_\text{dip}$, \cref{eq:E_dip}, which is an antiferromagnetic coupling. We note that the correlation between $\vb{m}$ and $\vb{u}$ increases with $\sigma$ but is completely independent of the dipole coupling for all $\lambda$. Conversely the correlations between $\vb{m}_1, \vb{m}_2$ and $\vb{r}$ depend only on $\lambda$. This demonstrates the statistical decoupling  of particle orientation and the other degrees of freedom mentioned above. From \cref{fig:force_comparison_stat_phys}b we observe that \cref{eq:Fdip_12} describes the force well, up to $\lambda \sim 1$.

\subsection{Van der Waals analogies}

\def\arraystretch{1.2}%
\begin{table}[tbh]
    \centering
    \begin{tabular}{|c|c|c|}
        \hline
        Interaction & Electric / Molecular & MNP analogue  \\
        \hline
        Keesom & Two polarised molecules & Two SMNPs\\
        \hline
        Debye & Polarised and unpolarised & SMNP and ZMC \\
        \hline
        Dispersion & Two unpolarised & Two ZMCs \\
        \hline
    \end{tabular}
    \caption{Contributions to the vdW interaction as described in Ref. \cite{israelachvili_intermolecular_2011} chapter 4-6, which combinations of molecules they apply to and analogous magnetic effects in MNP suspensions. SMNP stands for superparamagnetic nanoparticle and ZMC for zero-moment cluster, i.e.\ a compact assembly of MNPs with no net magnetisation in the groundstate.}
    \label{tab:vdW_analogy}
\end{table}

We note a strong conceptual similarity between \cref{eq:F_dip_avg} and the van der Waals (vdW) force, which originates from the correlated fluctuations of electric dipoles \cite{israelachvili_intermolecular_2011,bishop_nanoscale_2009}.
To be more precise, one may divide the vdW interaction into 3 types: the Keesom interaction between a pair of polar molecules, the Debye interaction between a polar molecule and the dipole it induces in a neutral molecule, and the dispersion interaction between neutral molecules due to temporary electric dipoles appearing because of quantum fluctuations (see e.g.\ chapters 4-6 of Ref. \cite{israelachvili_intermolecular_2011}). The greatest similarity is to the Keesom interaction, which is also derived by a thermal average over the orientation of permanent dipoles, but each effective potential has the same characteristic $1/r^6$ dependence as \cref{eq:averaged_attraction}. 

Analogously to the Debye interaction, if a lone MNP comes near an MNP cluster with zero net moment, the former induces a net magnetisation in the latter, yielding an attractive force. Two zero-moment clusters will also attract due to correlated, temporary magnetisation, which is analogous to the dispersion interaction except that the fluctuations are thermal rather than quantum.
In each case, there is an average force of attraction despite the magnetisation time-averaging to zero. Whether or not aggregation occurs depends if the attraction is large enough to overcome the translational, thermal fluctuations (Brownian motion).

\section{Discussion \label{sec:discussion}}


\subsection{When does anisotropy matter?}

We emphasize that because mechanical rotation is orders of magnitude slower than magnetic, $K$ will impact the timescale of thermal fluctuations and how quickly the MNPs react to magnetic perturbations. Thus for transient dynamics and how the system responds to time-varying stimuli like an external, alternating magnetic field, the anisotropy is important. One consequence is that $\sigma$ is irrelevant for the static susceptibility of a ferrofluid\cite{elfimova_static_2019}, but does affect hysteresis\cite{helbig_self-consistent_2023} and dynamic susceptibility\cite{ilg_nonequilibrium_2024}.

Since the simulations shown in \cref{fig:lifetime_simulations} had all dimers initiated in the groundstate rather than a thermal distribution, transient dynamics are included. However, the characteristic time for Brownian, rotational diffusion is\cite{einstein1906theory,ten_hagen_brownian_2011}
\begin{align*}
    \tau_B = \frac{\zeta_\text{r}}{2k_BT} = \frac{3\eta V}{k_BT},
\end{align*}
which is $0.4\:\mathrm{\mu s}$ for the default parameters. Hence at high $\lambda$ where magnetic attraction is significant, the magnetic moments reach a thermal distribution much faster than debonding can occur. Surface friction, lubrication effects\cite{russel_colloidal_1989,jeffrey_calculation_1984,jeffrey_calculation_1992,townsend_generating_2019} (the change in fluid behaviour when confined between the particle surfaces) or a more viscous medium will slow down this relaxation, but as long as the combined Néel-Brownian relaxation time\cite{shliomis_frequency_1993,stepanov_combined_1991} is short compared to the experiment, the equilibrium analysis remains valid.

A key assumption in the model is that the two MNPs in a dimer can rotate freely relative to each other even in surface contact. This is common in simulations\cite{durhuus_simulated_2021,satoh_brownian_1999,satoh_stokesian_1998,berkov_langevin_2006,chuan_lim_agglomeration_2012}, but in reality it depends on the nature of the aggregate and its surface chemistry. For example if electrostatic repulsion or a non-adhesive surfactant coating prevent the MNP cores from merging, they may retain their rotational freedom.
Conversely if the particles sinter or fuse together, or stick because of entangled surface polymers, they will move as a single rigid body. Then the magnetic moments can only reorganise by crossing anisotropy energy barriers, in which case anisotropy impacts the average magnetisation, thereby indirectly affecting aggregate stability and further aggregation. Analogously $K$ may alter the magnetisation of MNPs with non-uniform magnetisation and the resulting interactions.

That said, we have demonstrated in general that the phenomenon of superparamagnetism itself does not limit aggregation. Thermal fluctuations do decrease the effective force of attraction, favouring Brownian motion over aggregation, but whether the fluctuations are slow and mechanical or rapid moment reversals does not impact the average force nor the average aggregate lifetime. When considering the equilibrium behaviour of an MNP ensemble, e.g.\ the distribution of aggregates, what matters is 
the energy landscape, in particular energy barriers relative to $k_B T$.


In any case, the self-assembly of colloidal particles has to go through the dimer stage, and our results unambiguously indicate that anisotropy and associated phenomena do not influence this initial aggregation for uniformly magnetised particles, regardless of solvent or surface chemistry.

\subsection{Why are SMNP solutions easier to stabilise?}

We have found that SMNPs are just as prone to form dimers as blocked particles with equal $R, M$ and $T$, which at first glance appears to contradict the experimental observation that colloidal MNPs are easier to keep apart when superparamagnetic. However referring to \cref{eq:dimensionless_parameters,eq:lambda_surface_contact} we note that both the parameter governing superparamagnetism $\sigma$ and the one determining bond stability $\lambda(r=2R)$ are proportional to $\frac{R^3}{T}$. In fact
\begin{align}
    \frac{\lambda(r=2R)}{\sigma} = \frac{\mu_0 M^2}{24K},   \label{eq:lambda_sigma_ratio}
\end{align}
hence $\sigma$ and $\lambda$ can only be tuned independently via the material parameters $K, M$. For typical iron oxide parameters ($M=400\:\mathrm{kA/m}$ and $K = 20\:\mathrm{kJ/m^3}$) we find that $\sigma = 2.4\lambda$ so they are within an order of magnitude.

The characteristic times also have similar exponential dependencies, i.e.\ $\tau_\text{relax} \sim \ee^{\sigma}$ and $\tau_\text{debond} \sim \ee^{\lambda}$. For example if $R$ increases from 6 to 12 nm (default parameters otherwise), $\lambda$ goes from 1.8 to 14.6 which corresponds to a relative increase of ca.\ $\ee^{12.8} = 4 \cross 10^5$ in $\tau_\text{debond}$. Meanwhile $\tau_\text{relax}$ increases by 13 orders of magnitude. If we instead lower the temperature from 60 K to 30 K (default parameters otherwise), $\tau_\text{debond}$ increases by a factor 200 and $\tau_\text{relax}$ by $4\cross 10^5$. Here we neglect exponential prefactors and other details, like how $K$ is itself size dependent for nanoscale particles\cite{pisane_unusual_2017}, but the point remains: over a relatively small span of particle sizes and temperatures, Néel relaxation and debonding time both vary many orders of magnitude, and with the same qualitative trends, thus giving the false appearance of a connection.

For most MNP materials, we find that the ratio in \cref{eq:lambda_sigma_ratio} is close to unity or lower. FeNi particles are outliers though, as they can be synthesised to be magnetically soft while having a reasonably large saturation magnetisation. For instance Ref.\cite{kumari_investigating_2023} reports $K = 3.9\:\mathrm{kJ/m^3}$ and $M_s = 710\:\mathrm{kA/m}$ which implies $\lambda = 6.8 \sigma$. Thus FeNi-SMNPs are good candidates for observing magnetic aggregation of superparamagnets in zero-field conditions, with the characteristic linear structures this entails\cite{durhuus_simulated_2021}.



\section{Conclusions \label{sec:conclusions}}

We have demonstrated methods for computing the Néel relaxation time $\tau_\text{relax}$ for a rigid MNP dimer and the average debonding time $\tau_\text{debond}$ for an arbitrary colloidal dimer.
$\tau_\text{relax}$ quantifies the superparamagnetism, including how interactions exponentially suppress moment reversals, while $\tau_\text{debond}$ quantifies the strength of the magnetic bond and its stability against thermal fluctuations. Thus $\tau_\text{debond}$ can be used to estimate the tendency of colloids to aggregate and how said tendency scales with the relevant parameters.

While Néel relaxation is inherently magnetic, the notion of debonding time and the procedure to compute it is broadly applicable within colloid science. On one hand $\tau_\text{debond}$ is directly measurable, e.g.\ using electron microscopy, or optical microscopy for larger particles. On the other hand $\tau_\text{debond}$ can be used numerically to estimate bond stability, parameter dependence and characteristic timescales for arbitrary colloids, complementing more demanding many-body simulations.

Using these characteristic times, we have found unequivocally that the uniaxial anisotropy constant $K$ has no impact on aggregation within our model. Since varying $K$ interpolates from superparamagnetic to blocked particles, this implies that the degree of superparamagnetism does not affect aggregation. This holds both when $\tau_\text{relax}$ is close to and much longer than $\tau_\text{debond}$, so the explanation is not a matter of relative timescales. The reason for this surprising result is that even when the time-average magnetisation of every particle is zero, the thermally fluctuating magnetic moments are correlated and this yields a net attraction on average; a magnetic analogy of the vdW force. 

The key point is that we assume the individual MNPs are 
free to rotate mechanically. As all inanimate matter, the rotating magnetisation tends to take the path of least resistance, so if the MNP can skip the anisotropy barriers by rotating mechanically, then the magnetisation distribution decouples from anisotropy. This argument also holds for a many-particle suspension in a constant external field and for all higher order anisotropies. Whether the moment fluctuations are slow and mechanical (high $K$) or rapid, magnetic dynamics (low $K$) the average magnetic forces are the same as if $K=0$; and likewise for particle distribution and sample magnetisation in equilibrium. 

If several particles merge into a rigid cluster, or the individual MNP is not uniformly magnetised, then anisotropy may alter the magnetisation and resulting magnetostatic interactions. 
But the phenomenon of superparamagnetism, i.e.\ magnetisation time-averaging to zero due to moment-reversals, never affects aggregation.
The reason that SMNPs appear to aggregate less than blocked particles is that 
the parameters governing superparamagnetism and aggregation have the same $R^3/T$ dependence on particle size and temperature, hence for a given material the two are strongly correlated. 

The shift in conceptual understanding presented her, along with our auxiliary results on bonding and thermal fluctuations in dimers, is relevant not only in stabilisation of SMNP colloids, but also in controlled self-assembly\cite{hu_shaping_2019,berret_controlled_2006,krasia-christoforou_single-core_2020,kralj_magnetic_2015,nakata_chains_2008}, and for understanding correlations within MNP systems in general. 
We hope this will aid in the experimental interpretation and future modelling of magnetic nanoparticle suspensions.

\section*{Conflicts of interest}
There are no conflicts to declare.

\appendix

\section{Néel relaxation statistics \label{appsec:Néel_relaxation_statistics}}

For a Poisson distribution, the probability of $k$ events in time $t$ is
\begin{align*}
    P(k,t) = \left(\frac{t}{\tau_\text{ev}}\right)^k \frac{\ee^{-t/\tau_\text{ev}}}{k!}.
\end{align*}
where $\tau_\text{ev}$ is the average time between events. We define an event as the magnetic moment of a single MNP flipping from one energy minimum to the other, i.e. $\theta \approx 0 \xrightarrow{} \theta \approx \pi$ or vice versa.

Regarding the average magnetisation along $z$, a magnetic moment flipping an even number of times contributes $+1$ while an odd flip number contributes $-1$. It follows that
\begin{align*}
    \expval{m_z}(t) = \sum_{k=0}^\infty (-1)^k P(k, t) = \ee^{-t/\tau_\text{ev}} \sum_{k=0}^\infty \left(\frac{-t}{\tau_\text{ev}}\right)^k \frac{1}{k!} = \ee^{-2t/\tau_\text{ev}},
\end{align*}
where we used the Taylor expansion of $\ee^{-t/\tau_\text{ev}}$ for the last equality. With $\tau_\text{relax} = \tau_\text{ev}/2$ this yields \cref{eq:tau_relax}.

When the moment reaches the top of the anisotropy barrier ($\theta = \pi/2$) there is a 50/50 chance of falling back to the previous minimum rather than flipping. If we define an event as reaching this saddle point, then carry out the argument above the result is $\expval{m_z} = \ee^{-t/\tau_\text{ev}}$. Thus $\tau_\text{relax}$ equals the average time to reach the barrier top which is precisely half the average reversal time $\tau_\text{rev}$. Indeed, when tracking reversals numerically,
Kalmykov et.\ al.\cite{kalmykov_damping_2010} demonstrated that depending on the exact switching condition (e.g. $m_z < 0$ or $m_z < -0.9$), a factor of $1/2$ is required to reproduce the theoretical value of $\tau_\text{relax}$.

\section{Relaxation time in strong coupling limit \label{appsec:relaxation_time_in_strong_coupling_limit}}

The goal is to derive an analytical expression for the Néel relaxation time of a mechanically fixed dimer when the dipole-dipole coupling is strong (cf.\ \cref{sec:superparamagnetism_of_dimers}).

As exemplified in \cref{fig:tauRelax}a, when the coupling parameter $\lambda$ is high enough, the two moments tend to flip together and their fluctuations are strongly correlated. To make the problem tractable we assume the moments are perfectly correlated, so that knowing $\vb{m}_1$ yields $\vb{m}_2$ exactly. Then the magnetic dynamics are equivalent to those for a single MNP, except with an extra potential from the interaction 
as $\vb{m}_1$ essentially drags $\vb{m}_2$ along.

We make the additional assumption that when a reversal event happens, the moments flip along the path of least energy. That is, for every direction $\vb{m}_1$ points during a flip, the orientation of $\vb{m}_2$ minimises the system energy. By symmetry there is no reason for either moment to flip faster than the other, which entails equal polar angles, hence $\theta_1 = \theta_2 = \theta$ where $\cos \theta_i = m_{i,z}$. Then, expressing the dipole energy \cref{eq:E_dip} in spherical coordinates
\begin{align}
    E_\text{dip} = k_B T \lambda [\sin^2 \theta \cos (\phi_1 - \phi_2) - 2 \cos^2 \theta],
\end{align}
where $\phi_1$ and $\phi_2$ are azimuthal angles. The lowest energy occurs for $\phi_1 - \phi_2 = \pi$ in which case the total system energy from \cref{eq:system_energy} is
\begin{align}
    E = - k_B T (2 \sigma + \lambda) \cos^2 \theta
\end{align}
up to a constant.
Thus with perfectly correlated rotation along the lowest energy path, we get the same energy and consequent dynamics as in the single particle case, except that $\sigma$ is replaced by the effective anisotropy $2\sigma + \lambda$. Applying this substitution to the classic result by Brown\cite{brown_thermal_1963} yields \cref{eq:tau_relax_optimal_path}.

Browns proof starts by mapping the LLG equation \cref{eq:LLG} to the corresponding Fokker-Planck equation (FPE) (see e.g. Ref. \cite{aron_magnetization_2014} sec. 3.4 or the appendix of Ref. \cite{garanin_fokker-planck_1997} for more modern derivations of the FPE). The FPE is then written in spherical coordinates before applying the methods and approximations of Kramers' escape theory\cite{kramers_brownian_1940,hanggi_reaction-rate_1990}, which are valid in the high-barrier case $\Tilde{\sigma} \gg 1$ (see e.g.\ section IIIB of Ref. \cite{coffey_thermal_2012}). Since we assume $\lambda \gg 1$ already, Kramers' escape theory imposes no further restrictions. 
That said, our simulations indicate that uniaxial anisotropy favours individual, asynchronous reversals, so the present model is most accurate when $\lambda \gg \sigma$ (cf. \cref{fig:tauRelax}c).

\section{Dimer debonding time statistics\label{appsec:dimer_bonding_time_statistics}}

We defined the debonding time $\tau_\text{debond}$ from the exponential decay in unbroken bonds, cf.\ \cref{eq:diff_N_u}. Here we prove that $\tau_\text{debond}$ is also the average lifetime of a given dimer bond.

Let $P_u(t)$ be the probability that a given dimer is undivided at time $t$, i.e.\ bonded in the entire interval from 0 to $t$. From \cref{eq:diff_N_u} we have that
\begin{align*}
    P_u(t) = N_u(t)/N_0 = \ee^{-t/\tau_\text{debond}}.
\end{align*}
Now let $\rho(t) \dd t$ be the probability that the bond breaks precisely in the infinitesimal interval $\dd t$. Then
\begin{align*}
    P_u(t) = 1 - \int_0^t \rho(t') \dd t'.
\end{align*}
Differentiating wrt.\ $t$ the distribution of bond breaks over time is
\begin{align*}
    \rho(t) = -\dv{t} P_u(t) = \frac{1}{\tau_\text{debond}}\ee^{-t/\tau_\text{debond}}.
\end{align*}
The average lifetime is the expectation value of the bond breaking time, i.e.\
\begin{align}
    \int_0^\infty t \rho(t) \dd t = \frac{1}{\tau_\text{debond}} \int_0^\infty t \ee^{-t/\tau_\text{debond}} \dd t = \tau_\text{debond},
\end{align}
so the exponential decay time is also the average lifetime, as claimed.

\section{Correlation measures in weak coupling limit \label{appsec:correlation_measures_in_weak_coupling_limit}}

The goal is to derive correlations between the various degrees of freedom ($\vb{m}_i, \vb{u}_i, \vb{r}$) by calculating a number of ensemble averaged dot products, e.g.\ $\expval{\vb{m}_1 \vdot \vb{m}_2}$.

We consider the partition function \cref{eq:partition_function} for fixed $r$ and use spherical coordinates where $\vb{r} = r \vb{\hat{z}}$. With the shorthand $c_\theta = \cos \theta$ and $s_\theta = \sin \theta$, this amounts to
\begin{align*}
    \vb{m}_i = s_{\theta_i} c_{\phi_i}\vb{\hat{x}} + s_{\theta_i} s_{\phi_i}\vb{\hat{y}} +  c_{\theta_i}\vb{\hat{z}} 
\end{align*}
and
\begin{align*}
    \vb{m}_1 \vdot \vb{m}_2 = s_{\theta_1} s_{\theta_2} \underbrace{[c_{\phi_1} c_{\phi_2} + s_{\phi_1} s_{\phi_2}]}_{c_{\phi_2 - \phi_1}} + c_{\theta_1} c_{\theta_2},
\end{align*}
so, defining $\Delta \phi = \phi_2 - \phi_1$, the dipolar energy is
\begin{align*}
    E_\text{dip} = k_B T \lambda [s_{\theta_1} s_{\theta_2} c_{\Delta \phi} - 2 c_{\theta_1} c_{\theta_2}],
\end{align*}
and the partition function becomes
\begin{align*}
    Z &= I_0^2 \int \dd\vb{m}_1 \dd\vb{m}_2 \: \ee^{- E_\text{dip}/k_BT} 
    \\
    &= I_0^2 \int_0^{2\pi} \dd\phi_1\dd\phi_2 \int_0^\pi \dd\theta_1\dd\theta_2 \: s_{\theta_1} s_{\theta_2} \ee^{\lambda (2c_{\theta_1}c_{\theta_2} - c_{\Delta \phi}s_{\theta_1}s_{\theta_2})}.
\end{align*}
Taylor expanding in $\lambda$, we note that for all odd powers of $\lambda$ the integrand changes sign under $\vb{m}_1 \xrightarrow{} -\vb{m}_1$ hence integrates to zero. This leaves
\begin{align*}
    Z = 16 \pi^2 I_0^2 + \order{\lambda^2}.
\end{align*}
In $\expval{\vb{m}_1 \vdot \vb{m}_2}$, the same anti-symmetry under inversion makes even powers integrate to zero. Thus
\begin{align}
    \expval{\vb{m}_1 \vdot \vb{m}_2} &= -\frac{2\pi I_0^2}{Z} \int_0^{2\pi} \dd\Delta\phi \int_0^\pi \dd\theta_1 \dd\theta_2\: s_{\theta_1}s_{\theta_2} (\vb{m}_1 \vdot \vb{m}_2) \frac{E_\text{dip}}{k_B T} + \order{\lambda^3} 
    \notag \\
    &= \frac{2\pi I_0^2 \lambda}{Z} \int_0^\pi \dd\theta_1 \dd\theta_2 s_{\theta_1} s_{\theta_2} (4\pi c_{\theta_1}^2 c_{\theta_2}^2 - \pi s_{\theta}^2 s_{\theta_2}^2) + \order{\lambda^3}
    \notag \\
    &= \order{\lambda^3}.   \label{eq:<m1.m2>}
\end{align}
The third order term is non-zero, but we only evaluate up to second order here. 

From the dipolar energy \cref{eq:E_dip}, we expect the product $(\vb{m}_1 \vdot \vb{\hat{r}})(\vb{m}_2 \vdot \vb{\hat{r}})$ to be positive, i.e.\ for the two magnetic moments to align along the interparticle axis. Indeed a calculation similar to \cref{eq:<m1.m2>} yields
\begin{align}
    \expval{(\vb{m}_1 \vdot \vb{\hat{r}})(\vb{m}_2 \vdot \vb{\hat{r}})} 
    &= \frac{2}{9}\lambda + \order{\lambda^3}.  \label{eq:parallel_correlation}
\end{align}

For completeness, we also consider the normalised component perpendicular to $\vb{r}$:
\begin{align*}
    \vb{m}_\perp 
    = c_{\phi} \vb{\hat{x}} + s_{\phi} \vb{\hat{y}}.
\end{align*}
The dot product is simply $\vb{m}_{1\perp} \vdot \vb{m}_{2\perp} = c_{\Delta \phi}$, which averages to
\begin{align}
    \expval{\vb{m}_{1\perp} \vdot \vb{m}_{2\perp}} 
    &= -\frac{\pi^2}{32}\lambda + \order{\lambda^3}.    \label{eq:transverse_correlation}
\end{align}
The negative value indicates that the transverse moments tend to point anti-parallel. The combination of ferro- and antiferromagnetic coupling inherent in $E_\text{dip}$ and expressed in \cref{eq:parallel_correlation,eq:transverse_correlation}, explains how the linear term in $\vb{m}_1 \vdot \vb{m}_2$ can average to zero.

Finally, for a magnetic moment and the corresponding anisotropy axis we find that $\expval{\vb{m}_i \vdot \vb{u}_i} = 0$ and
\begin{align*}
    \expval{(\vb{m}_i \vdot \vb{u}_i)^2} &= \frac{1}{Z} \int \dd\vb{m}_i \dd\vb{m}_2 \ee^{-E_\text{dip}/k_BT} \int (\vb{m}_i \vdot \vb{u}_i)^2 \ee^{\sigma (\vb{m}_i \vdot \vb{u}_i)^2} \dd \vb{u}_i
    \\
    &= \frac{\int (\vb{m}_i \vdot \vb{u}_i)^2 \ee^{\sigma (\vb{m}_i \vdot \vb{u}_i)^2} \dd \vb{u}_i}{\int \ee^{\sigma (\vb{m}_i \vdot \vb{u}_i)^2} \dd \vb{u}_i}.
\end{align*}
Using a coordinate system where $\vb{m}_i = \vb{\hat{z}}$ and spherical coordinates
\begin{align}
    \expval{(\vb{m}_i \vdot \vb{u}_i)^2} = \frac{\int_0^\pi c_\theta^2 s_\theta \ee^{\sigma c_\theta^2} \dd\theta}{\int_0^\pi s_\theta \ee^{\sigma c_\theta^2} \dd\theta} = \frac{\xi'}{\xi},   \label{eq:<u.m>}
\end{align}
where
\begin{align*}
    \xi=\int_{-1}^1 \ee^{\sigma x^2}\dd x \qq{and} \xi' = \dv{\sigma}\xi = \int_{-1}^1 x^2 \ee^{\sigma x^2} \dd x.
\end{align*}
\Cref{eq:<u.m>} holds for a many-particle ferrofluid with interactions and does not use any perturbative approximations; only the assumption that each MNP can freely rotate mechanically.

\section{Orientation averaged dipole force \label{appsec:effective_dipole_force}}

Here we derive an effective interaction force and potential by averaging the dipole-dipole force over a thermal distribution of particle orientations $\vb{u}$ and moment directions $\vb{m}$. 

The term $\expval{(\vb{m}_1 \vdot \vb{m}_2) \vb{\hat{r}}}$ is zero to second order in $\lambda$ (cf. \cref{eq:<m1.m2>}). For the other components of the dipole force \cref{eq:Fdip_12} it follows from rotation symmetry around $\vb{\hat{z}}$, and \cref{eq:parallel_correlation} that
\begin{align*}
    \expval{(\vb{m}_1 \vdot \vb{\hat{r}})\vb{m}_2} = \expval{(\vb{m}_2 \vdot \vb{\hat{r}})\vb{m}_1} = \expval{(\vb{m}_1 \vdot \vb{\hat{r}})(\vb{m}_2 \vdot \vb{\hat{r}})} \vb{\hat{r}} = \frac{2}{9}\lambda \vb{\hat{r}},
\end{align*}
to second order. Inserting in \cref{eq:Fdip_12} we get \cref{eq:F_dip_avg}. Interestingly, the leading order attraction is exclusively due to $\expval{(\vb{m}_1 \vdot \vb{\hat{r}})(\vb{m}_2 \vdot \vb{\hat{r}})}$, i.e.\ the tendency of the magnetic moments to align along the interparticle axis.

\section{Numerical ensemble average \label{appsec:numerical_ensemble_average}}

To average out the thermal noise and compare simulations with statistical physics results, it is useful to consider ensemble averages, for example the average dipole force between MNP pairs at a given distance $\expval{\vb{F}_\text{dip}}(r)$.

To simulate this, we consider the somewhat unphysical case where the relative positions of two MNPs are fixed (constant $r$), but they rotate both magnetically and mechanically, i.e.\ $\vb{u}_1,\vb{u}_2,\vb{m}_1$ and $\vb{m}_2$ vary according to \cref{eq:LLG,eq:rotation}.
Because a given MNP pair will cycle through the full orientation space on a timescale we can simulate, time- and ensemble averages are equivalent (the ergodic hypothesis holds). For convenience we use a combined average over time and $N_0$ MNP pairs. As elsewhere, we initiate with all vectors aligned as in \cref{fig:Dimer_system_sketch}b, however we give the system a time $t_\text{eq}$ to reach thermal equilibrium before averaging. Thus for a given quantity $Q$ we define the numerical ensemble average by:
\begin{align*}
    \expval{Q} = \frac{1}{N_0}\sum_{i\in \text{pairs}} \frac{1}{t_\text{sim} - t_\text{eq}} \int_{t_0}^{t_\text{sim}} Q_i \dd t,
\end{align*} 
where $t_\text{sim}$ is the simulation time.

$N_0 = 250$ in all simulations. For $t_\text{eq}$ we use twice the characteristic time for Brownian, rotational diffusion\cite{einstein1906theory,ten_hagen_brownian_2011}:
\begin{align*}
    t_\text{eq} = \frac{\zeta_\text{r}}{2k_BT} = \frac{3\eta V}{k_BT},
\end{align*}
which is $0.8\:\mathrm{\mu s}$ for the default parameters (see \cref{tab:parameters}).

\newpage

\bibliography{references.bib}

\bibliographystyle{rsc} 

\end{document}


\title{\textbf{Supplementary information for the article \\ ``Dipolar Attraction of Superparamagnetic Nanoparticles''}}
\author{Frederik L. Durhuus}
\affiliation{Department of Physics, Technical University of Denmark, 2800 Kgs.\ Lyngby, Denmark}
\author{Marco Beleggia}
\affiliation{Department of Physics, University of Modena and Reggio Emilia, 41125 Modena, Italy}
\affiliation{DTU Nanolab, Technical University of Denmark, 2800 Kgs. Lyngby, Denmark}
\author{Cathrine Frandsen}
\affiliation{Department of Physics, Technical University of Denmark, 2800 Kgs.\ Lyngby, Denmark}
\email{fraca@fysik.dtu.dk}

\maketitle

Note that for all simulations in the supplementary material, we use default parameters unless otherwise indicated (cf.\ Tab. 1 of the main text).

\section{Convergence tests}

\begin{figure}
    \centering
    \includegraphics[width=\columnwidth]{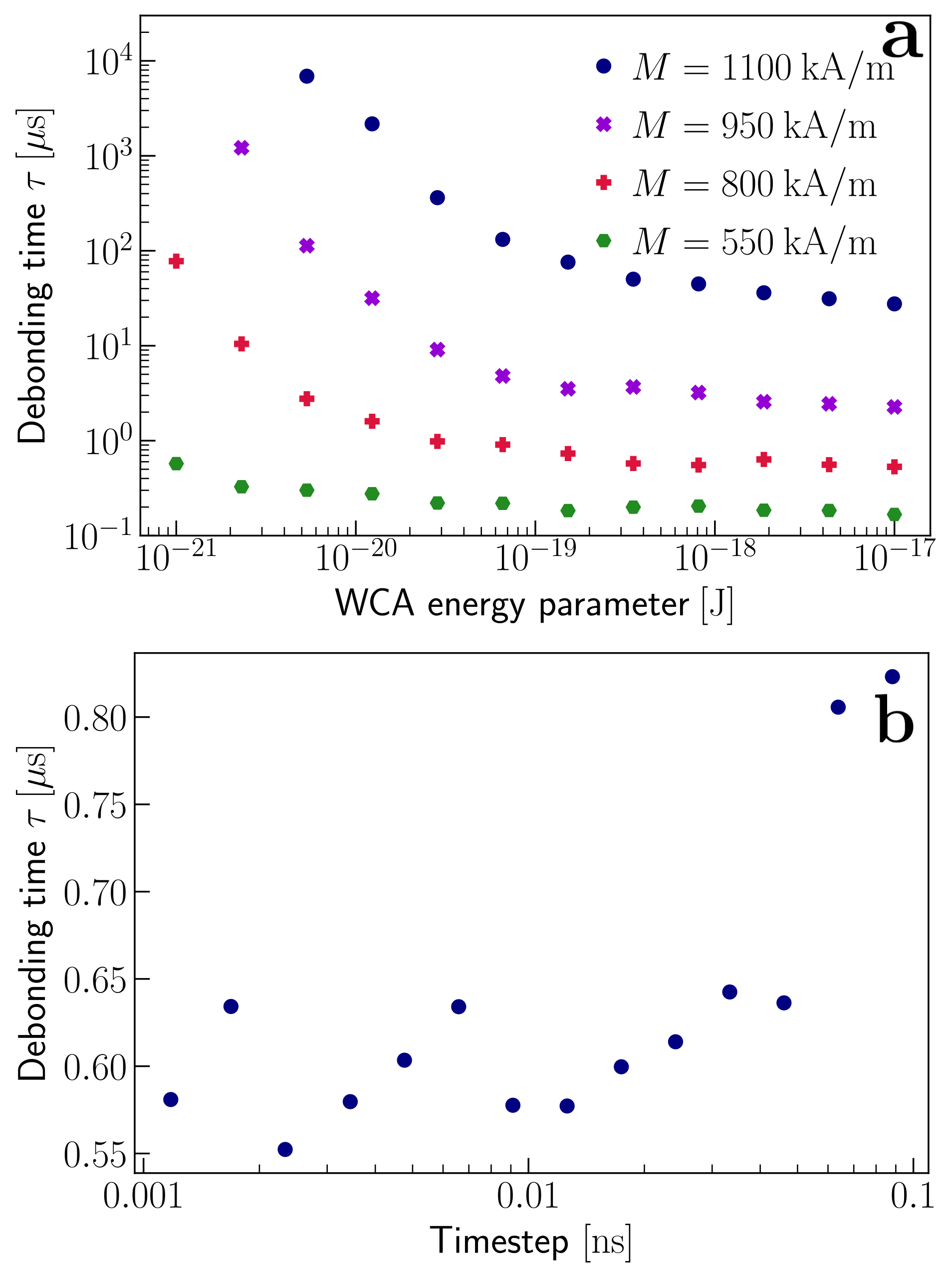}
    \caption{\textbf{a)} Debonding time vs.\ WCA energy parameter for four different magnetisations. \textbf{b)} Debonding time vs.\ numerical timestep}
    \label{fig:convergence_tests}
\end{figure}

As noted in the main paper, we are primarily interested in magnetic interactions and how they are affected by thermal fluctuations, but we need an additional interaction to model the solidity of the MNP surface. To implement this steric repulsion, we use the Week-Chandler-Anderson force:
\begin{align}
    \vb{F}_\text{WCA} = \begin{cases}
    12 \epsilon_\text{WCA} \left[\frac{(2R)^{12}}{r^{13}} -  \frac{(2R)^6}{r^7}\right] 
    & r \leq 2R
    \\
    0 & r > 2R
    \end{cases} 
    \label{eq:F_WCA}
\end{align}
where we recall that $R$ is particle radius, $r$ is center-to-center separation and $\epsilon_\text{WCA}$ is the WCA energy parameter. Ideally we would let $\epsilon_\text{WCA} \xrightarrow{} \infty$ to get a hard-sphere model, but this is numerically unstable. Thus we need $\epsilon_\text{WCA}$ low enough for stable collisions, but high enough so the slight MNP overlap does not significantly affect the end results. 

In \cref{fig:convergence_tests}a we present a convergence study in $\epsilon_\text{WCA}$. We see that a higher magnetisation necessitates a stronger WCA repulsion, because of the increased magnetostatic attraction. That said, $\epsilon_\text{WCA} = 10^{-17} \:\mathrm{J}$ is sufficient for all simulated $M$ values, so this is the value we used. $10^{-18}\:\mathrm{J}$ also looks adequate, but when we include the LLG equation, magnetic dynamics requires a very short timestep compared to the timescale of translational motion, so we can afford to use a very hard potential.

In \cref{fig:convergence_tests}b we show a convergence test in the timestep used for our numerical integration of the governing equations (see appendix of Ref. \cite{durhuus_conservation_2024} for the algorithm we use). While there is some noise due to the stochastic nature of $\tau$, we see that the $0.01\:\mathrm{ns}$ timestep we simulated with is well converged.

\section{Effect of bonding criterion}

\begin{figure}[tbh]
    \centering
    \includegraphics[width=\columnwidth]{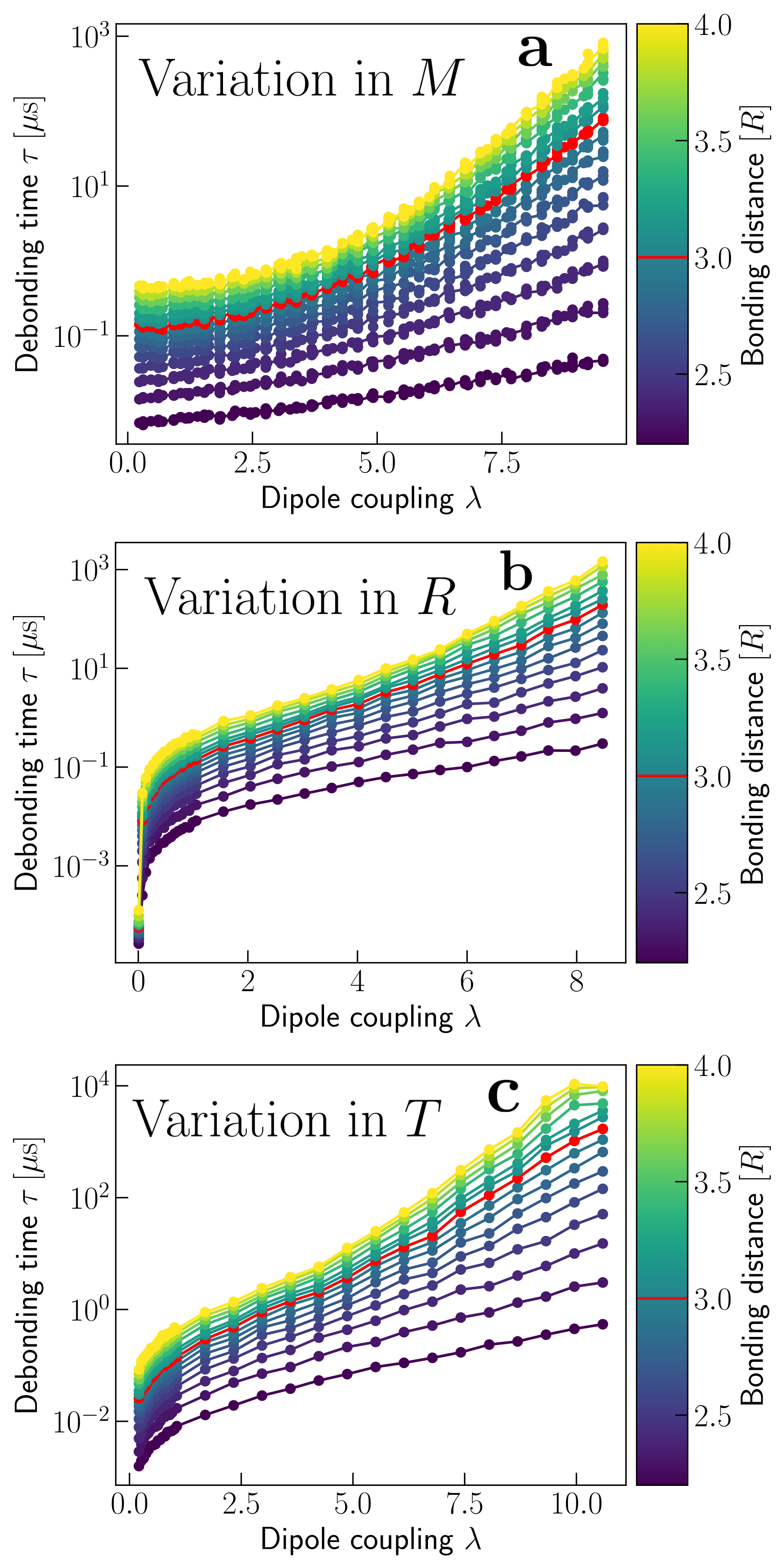}
    \caption{Debonding time vs.\ dipole coupling strength and critical bonding distance. $\lambda$ is varied via magnetisation in \textbf{a)}, particle radius in \textbf{b)} and temperature in \textbf{c)}. The bonding distance in units of particle radius is colourcoded, with the data shown in Fig.\ 4 of the main text coloured red.}
    \label{fig:dist_bond_effect}
\end{figure}

In the main text, we define debonding time as the instant the pair distance $r$ exceeds the critical value $r_\text{bond} = 3R$. The choice of $r_\text{bond}$ is somewhat arbitrary, so in \cref{fig:dist_bond_effect} we vary the bonding criterion for the same dataset as in Fig.\ 4 of the main text.

We see that when $r_\text{bond}$ is low, the $\tau(\lambda)$ curve flattens out because even when strongly coupled $r$ often exceeds $r_\text{bond}$ briefly due to translational Brownian motion. For higher $r_\text{bond}$ the curves are consistent in shape and more closely spaced, but still shifted increasingly upwards. Hence we can infer the parameter scaling and qualitative behaviour of $\tau$ as done in the main paper, but the exact value is a function of $r_\text{bond}$, which is currently ambiguous.

\section{Effect of viscosity on debonding time}

\begin{figure}[htb]
    \centering
    \includegraphics[width=\columnwidth]{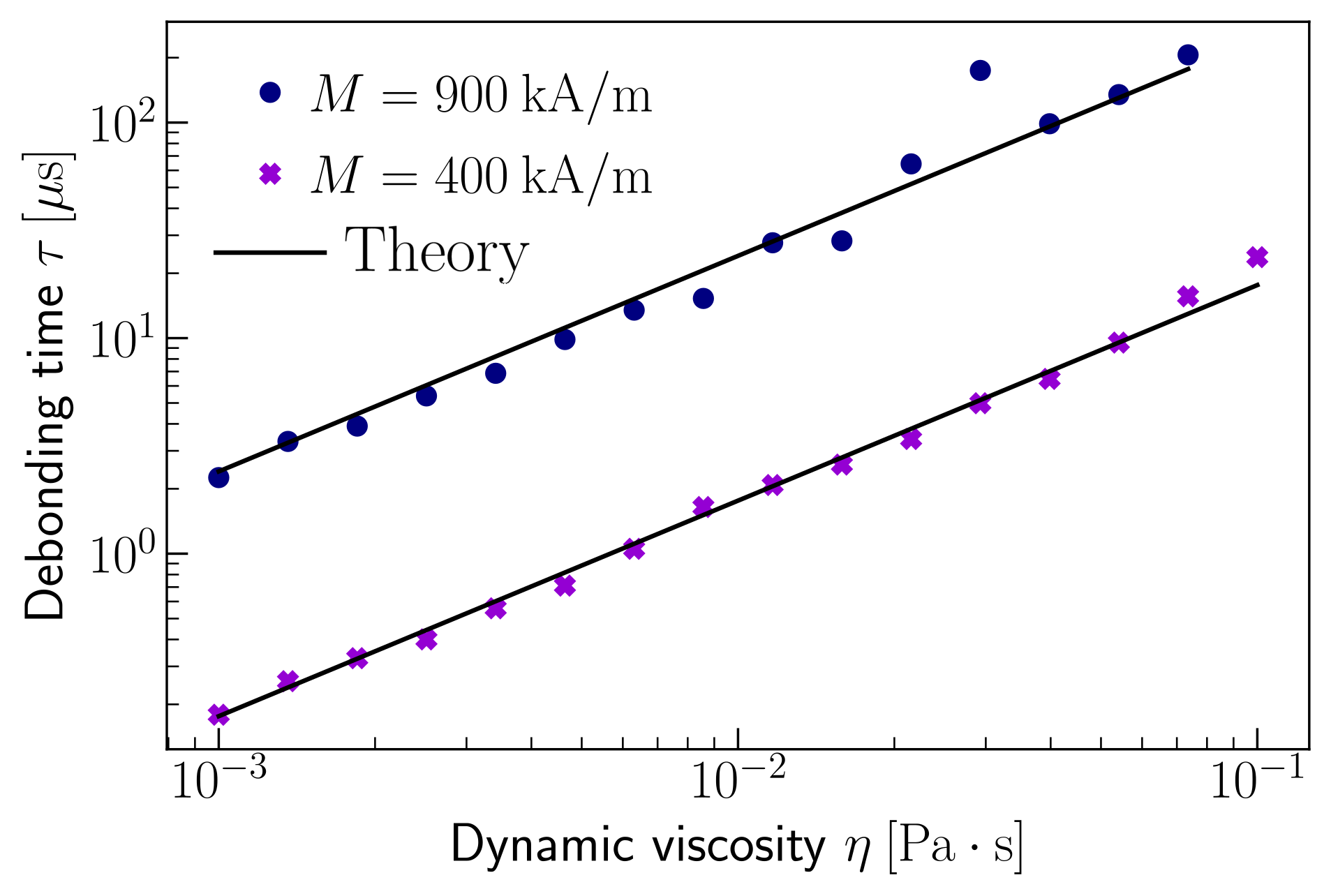}
    \caption{Debonding time vs. viscosity for two magnetisation values. The theory curves are fits to $\ln \tau = A + \ln \eta$ where $A$ is the only free parameter.}
    \label{fig:tau_vs_eta}
\end{figure}

Using dimensional analysis we inferred that the debonding time must be directly proportional to viscosity. In \cref{fig:tau_vs_eta} we verify this by plotting $\tau(\eta)$ on a double logarithmic scale and fitting to $\ln \tau = A + \ln \eta$. Since $A$ is the only free parameter, we are not tuning the slope of the line. The excellent agreement between simulation and theory proves that $\ln \tau = A + \ln \eta \xRightarrow{} \tau = \eta \ee^A$,
for both weak and strong dipole coupling.

\bibliography{references.bib}